\documentclass[aps,prd,preprint,groupedaddress,12pt,a4paper,superscriptaddress]{revtex4-2}

\usepackage{amsmath,amssymb,amsfonts}
\usepackage{amsthm}
\usepackage{graphicx}
\usepackage{multirow}
\usepackage{array}
\usepackage{booktabs}
\usepackage{mathrsfs}
\usepackage{textgreek}
\usepackage{algorithm}
\usepackage{algorithmicx}
\usepackage{algpseudocode}
\usepackage{float}
\usepackage{subcaption}
\usepackage{xcolor}
\usepackage{textcomp}
\usepackage{listings}
\usepackage{lipsum}
\usepackage[utf8]{inputenc}

\usepackage{booktabs}
\usepackage{multirow}

\begin{document}

\title{Exclusive process $\gamma \gamma \rightarrow J/\psi+\gamma$ production in ultraperipheral proton and nuclear collisions at the HL-LHC and FCC}

\author{Meng-Kun Jia}
\affiliation{School of Physics, Liaoning University, Shenyang 110036, China}

\author{Xiao-Bo Jin}
\affiliation{Department of Physics and Electronics, School of Mathematics and Physics, Beijing University of Chemical Technology, Beijing 100029, China}

\author{Kui-Yong Liu}
\email{liukuiyong@lnu.edu.cn}
\affiliation{School of Physics and Electronic Technology, Liaoning Normal University, Dalian 116029, China}
\affiliation{School of Physics, Liaoning University, Shenyang 110036, China}

\author{Guang-Zhi Xu}
\email{xuguangzhi@lnu.edu.cn}
\affiliation{School of Physics, Liaoning University, Shenyang 110036, China}

\date{}

\begin{abstract}
We present a next-to-leading-order (NLO) analysis of exclusive $J/\psi+\gamma$ production via photon–photon fusion in ultraperipheral collisions at the High-Luminosity Large Hadron Collider (HL-LHC) and the Future Circular Collider (FCC). The study is performed within the NRQCD factorization framework for proton–proton, proton–nucleus, and nucleus–nucleus collisions, with nuclear species spanning a wide range of nuclear charges (O, Ca, Ar, Kr, Xe, and Pb), enabling a systematic investigation of nuclear effects on the production cross sections. The photon fluxes are modeled using an electric-dipole form factor, with the impact-parameter dependence strictly enforced to ensure the exclusivity of the process. We present predictions for total cross sections and kinematic distributions at leading order and NLO. 
With a transverse momentum cut of $p_T > 2\,\mathrm{GeV}$, the NLO corrections reduce the cross section by approximately $35\%$ at the central scale. The associated theoretical uncertainties are systematically estimated via renormalization scale variations.
Despite this suppression, the cross sections remain sizable and indicate that exclusive $J/\psi + \gamma$ production serves as a sensitive probe of photon-induced quarkonium production mechanisms. We thus present the corresponding event yields predicted for the HL-LHC and the FCC.
\end{abstract}


\maketitle

\section{Introduction}\label{sec1}

Charmonium production provides an important testing ground for quantum chromodynamics (QCD), as it involves both perturbative heavy-quark production and nonperturbative quarkonium formation. Among the charmonium states, the $J/\psi$ meson plays a central role owing to its clean experimental signature and relatively narrow width. Measurements of $J/\psi$ production in hadronic collisions over a wide range of energies have yielded extensive information on heavy-quarkonium production mechanisms and remain a key probe of QCD dynamics~\cite{LHCb:2015foc,CDF:2009kwm,PHENIX:2009ghc,Faccioli:2022alj,Xu:2012am,Li:2019mdx}.

The theoretical description of heavy quarkonium production is commonly formulated within the nonrelativistic QCD (NRQCD) factorization framework~\cite{Bodwin:1994jh}. This approach exploits the hierarchy of energy scales in quarkonium systems to separate short-distance coefficients (SDCs) from long-distance matrix elements (LDMEs). A characteristic feature of NRQCD is the inclusion of color-octet contributions, which allows a systematic expansion in the heavy-quark velocity and provides a unified description of quarkonium production in different reactions.

Most studies of $J/\psi$ production have concentrated on hadronic collisions dominated by strong interactions, where multiple partonic subprocesses and soft interactions obscure the interpretation of the production mechanism. Photon-induced reactions provide a complementary environment in which the production dynamics can be examined with substantially reduced hadronic backgrounds. In such processes, the interaction occurs at large impact parameters where the hadronic overlap is negligible and electromagnetic interactions dominate, a characteristic feature of ultraperipheral collisions (UPCs)~\cite{Baltz:2007kq,Bertulani:2005ru,Klein:2019qfb,Lansberg:2024zap}. In particular, photon–photon interactions within these UPCs offer a clean channel to probe quarkonium production driven primarily by electromagnetic fields~\cite{Klein:2020fmr}.

UPCs at the LHC realize this regime experimentally, enabling photon-induced reactions to be studied at unprecedented energies~\cite{Klein:2019qfb}. In UPC events, the colliding hadrons or nuclei pass each other with impact parameters larger than the sum of their radii, suppressing hadronic interactions while allowing their Lorentz-boosted electromagnetic fields to interact. These strong electromagnetic fields can be described in terms of fluxes of quasireal photons within the equivalent photon approximation (EPA)~\cite{Brodsky:1971ud,Budnev:1975poe}, effectively turning hadron colliders into powerful photon–photon and photon–hadron colliders.

Extensive measurements of exclusive $J/\psi$ production in UPCs have been reported across various collision systems, including $pp$ at LHCb~\cite{LHCb:2013nqs,LHCb:2014acg,LHCb:2016oce}, p--Pb and Pb--Pb at ALICE~\cite{ALICE:2013wjo,ALICE:2018oyo,ALICE:2023mfc,ALICE:2023kgv}, and Au--Au at PHENIX~\cite{PHENIX:2009xtn}. These results demonstrate that photon-induced quarkonium production can be studied with high precision, establishing UPCs as an ideal laboratory for investigating the underlying production mechanisms.

On the theory side, charmonium production in UPCs has been studied within various phenomenological approaches, including photon–nucleus and photon–photon induced mechanisms~\cite{Rapp:2008tf,Klasen:2008mh,Zheng:2024mep,Jiang:2024vuq,Obertova:2024nmb,Azevedo:2024bqd}. Inclusive charmonium production in photon–photon collisions has been investigated at NLO in direct photoproduction~\cite{Klasen:2004az}, and tree-level studies of $\gamma\gamma \to J/\psi + X$ in UPCs were performed using the \textsc{MadOnia} framework ~\cite{Klasen:2008mh}. More recently, increasingly exclusive final states have attracted attention. Exclusive $J/\psi$ plus jet production in ultraperipheral Pb–Pb collisions was analyzed at leading order within NRQCD in Ref.~\cite{Goncalves:2023sts}. A dedicated NRQCD study of exclusive $\gamma\gamma \to J/\psi + \gamma$ production in ultraperipheral Pb–Pb and $pp$ collisions, including corrections up to $\mathcal{O}(\alpha_s v^2)$, was presented in Ref.~\cite{Chen:2025gwp}, demonstrating the importance of higher-order effects.

In this work, we investigate exclusive $J/\psi+\gamma$ production via photon–photon fusion in UPCs within the NRQCD factorization framework. The analysis spans a wide range of collision systems, including proton–proton (pp), proton–nucleus (p--A), and nucleus–nucleus (A--A) interactions at the HL-LHC, as well as $pp$, p--Pb, and Pb--Pb collisions at FCC. The elastic photon flux of the proton and nuclei is modeled using the electric-dipole form-factor (EDFF) parametrization~\cite{Shao:2022cly}. The impact-parameter $b$ dependence is included explicitly, and the UPCs is implemented through the no-inelastic-interaction probability $P_{\mathrm{noinel}}\!\left(|\mathbf{b}_1-\mathbf{b}_2|\right)$, which acts as a hadronic survival factor suppressing configurations with hadronic overlap. We provide predictions for total cross sections and for transverse-momentum ($p_T$) and rapidity ($y$) distributions at both LO and NLO. Compared with previous studies, this work significantly broadens the phenomenological coverage of this channel by providing a unified treatment across multiple collider systems and nuclear species.

This paper is organized as follows. In Sec.~\ref{sec2}, we describe the theoretical framework underlying the NLO calculation, including the NRQCD factorization formalism, the treatment of photon fluxes with explicit impact-parameter dependence, and the enforcement of the exclusivity condition via the suppression of inelastic hadronic interactions in UPCs. Numerical predictions for total cross sections, differential distributions, and event yields are presented in Sec.~\ref{sec3}. Finally, Sec.~\ref{sec4} contains a summary of our main findings and conclusions.

\section{Theoretical Framework}
\label{sec2}
Under the NRQCD framework, the differential cross section of the exclusive process via photon fusion in an UPC of hadrons A and B with charges $Z_{1,2}$ can be written as:
\begin{equation}
	\label{fra:1}
	\sigma(AB \rightarrow J/ \psi +\gamma) =  \sum_{n}\int \frac{\mathrm{d} E_{\gamma_{1}}}{E_{\gamma_{1}}} \frac{\mathrm{d} E_{\gamma_{2}}}{E_{\gamma_{2}}} \frac{\mathrm{d}^{2} N_{\gamma_{1}/Z_{1}, \gamma_{2}/Z_{2}}^{(\mathrm{AB})}}{\mathrm{d} E_{\gamma_{1}} \mathrm{d} E_{\gamma_{2}}}
	\times \mathrm{d}\hat{\sigma}(\gamma_1 \gamma_2 \to c\bar{c}[n] + \gamma) \otimes \langle 0 | \mathcal{O}_{n}^{J/\psi} | 0 \rangle
\end{equation}
the SDCs $\mathrm{d}\hat{\sigma}(\gamma_1 \gamma_2 \to c\bar{c}[n] + \gamma) $ creating a charm quark pair ($c\bar{c}$) and an extra photon is given by:
\begin{equation}
	\label{fra:2}
	\mathrm{d}\hat{\sigma}(\gamma_1 \gamma_2 \to c\bar{c}[n] + \gamma)
	= \frac{1}{8\pi} \frac{| \overline{M} |^{2}}{2\hat{s}} 
	\, \mathrm{d}p_{T}^{2}\,\mathrm{d}y \,,
\end{equation}
where the index $[n]$ labels the color, spin, and angular-momentum quantum numbers of the produced $c\bar{c}$ pair. Here, $s$ denotes the total center-of-mass energy squared of the hadronic collision system, and $\hat{s} = x_1 x_2 s$ is the center-of-mass energy squared of the initial-state photons in the partonic frame, with  $x_i$ representing the parton momentum fraction carried by one of the initial photons, $x_i=E_{\gamma_i}/E_{beam}$. $p_T$ denotes the transverse momentum of the $c\bar{c}$ pair and $y$ is its rapidity. The LDMEs $\langle 0 | \mathcal{O}_{n}^{J/\psi} | 0 \rangle$ quantify the nonperturbative probabilities for a $c\bar{c}$ pair in a given Fock state $n$ to hadronize into $J/\psi$. In this paper, we only consider the color-singlet channel. The color-singlet LDME is related to the squared radial wave function at the origin \cite{Eichten:1995ch},
\begin{equation}
	\label{fra:3}  
	\langle 0 | \mathcal{O}^{J/\psi}(^3S_1^{[1]}) | 0 \rangle=\frac{6N_c}{4\pi}|R_{S}(0)|^2.
\end{equation}

In studies of $\gamma\gamma$ fusion in UPCs, the finite size of the hadrons implies that hadronic interactions may occur at small impact parameters, thereby spoiling the exclusivity of the process. To suppress such contributions, a geometric exclusion procedure is commonly adopted. In practice, this is implemented by introducing a minimum impact parameter ($b_{\text{min}} = R_{A,B}$ ) in the photon flux, together with an effective constraint equivalent to the geometric condition $|\mathbf{b}_1 - \mathbf{b}_2| > R_A + R_B$ \cite{Cahn:1990jk}. In Eq. \eqref{fra:1}, the component $\mathrm{d}^{2} N_{\gamma_{1}/Z_{1}, \gamma_{2}/Z_{2}}^{( \mathrm{AB} )}/(\mathrm{d} E_{\gamma_{1}} \mathrm{d} E_{\gamma_{2}})$ corresponds to the expressions as follows,

\begin{equation}
	\label{fra:4}   
	\begin{aligned}		
		\frac{\mathrm{d}^{2} N_{\gamma_{1}/Z_{1}, \gamma_{2}/Z_{2}}^{( \mathrm{AB} )}}{\mathrm{d} E_{\gamma_{1}} \mathrm{d} E_{\gamma_{2}}} 
		= \int & \mathrm{d}^{2}\mathbf{b}_{1} \mathrm{d}^{2}\mathbf{b}_{2}  P_{\text{noinel}}(|\mathbf{b}_1 - \mathbf{b}_2|) \\
		& \times N_{\gamma_{1}/Z_{1}}(E_{\gamma_{1}},\mathbf{b}_1) N_{\gamma_{2}/Z_{2}}(E_{\gamma_{2}},\mathbf{b}_2) \\
		& \times \theta(b_1 - \epsilon R_A) \theta(b_2 - \epsilon R_B),
	\end{aligned}
\end{equation}
It is formulated via the convolution of the two photon number densities $N_{\gamma_{i}/Z}(E_{\gamma_{i}},\mathbf{b}_{i})$, which are functions of the photon energies $E_{\gamma_{1,2}}$ and $\mathbf{b}_1$ is the vector from the center of particle A to the effective emission point of its associated photon. Similarly, $\mathbf{b}_2$ is defined for particle B. $R_{\mathrm{AB}} $ is effective charge radius of the two photon. The probability $P_{\mathrm{noinel}}(|\mathbf{b}_1 - \mathbf{b}_2|)$ for the absence of inelastic hadronic interactions is governed by conventional opacity or eikonal expressions \cite{Glauber:1970jm}, as expressed in Eq.~\eqref{fra:p}:

\begin{equation}
	\label{fra:p}  
	P_{\mathrm{noinel}}(\mathbf{b}) = 
	\begin{cases} 
		1 - e^{- \sigma^{NN}_{inel} T_{\text{AB}}(\mathbf{b})}, & \text{for nucleus-nucleus UPCs} \\
		1 - e^{- \sigma^{NN}_{inel} T_{\text{A}}(\mathbf{b})}, & \text{for proton-nucleus UPCs} \\
		1 - \Gamma(s_{NN}, \mathbf{b}) \quad \text{with} \quad \Gamma(s_{NN}, b) \propto e^{-b^2/(b_0)}, & \text{for proton-proton UPCs}
	\end{cases}
\end{equation}
the functions $T_{\text{A}}(\mathbf{b})$ and $T_{\text{AB}}(\mathbf{b})$ represent the nuclear thickness and overlap, respectively. These values of theirs are usually obtained by calculating the transverse density distribution of hadrons \cite{Loizides:2017ack,dEnterria:2020dwq}. The corresponding expression reads: 
\begin{align}
	\label{6}
	&T_A(\mathbf{b}) = \int \rho_A(\mathbf{b}, z) \, \mathrm{d}z, \\
	&T_{AB}(\mathbf{b}) = \int \mathrm{d}^2\mathbf{b}_1\, T_A(\mathbf{b}_1) \, T_B(\mathbf{b}_1 - \mathbf{b}),
\end{align}
 where the nuclear density distribution $\rho_A(\mathbf{b}, z)$ is commonly described by a generalized Woods-Saxon (Fermi) profile, which incorporates three essential parameters \cite{Loizides:2014vua}:

\begin{equation}
	\label{eq:fermi3}
	\rho_A(r) = \rho_0 \, \frac{1 + w_A\left(r/R_A\right)^2}
	{1 + \exp\!\left(\dfrac{r-R_A}{a_A}\right)},
\end{equation}
where $r = \sqrt{b^2 + z^2}$, $R_A$ is the nuclear radius, $a_A$ is the surface diffuseness, and $w_A$ is a deformation parameter. From the normalization condition:
\begin{equation}
	\int \mathrm{d}^3r\,\rho_A(r)=A,
\end{equation}
we can get normalization constant $\rho_0$, and $A$ is the mass number. The inelastic $NN$ cross section $\sigma^{NN}_{inel}$ is parameterized as a function of $\sqrt{S_{NN}}$, following the prescription in Ref. \cite{Loizides:2017ack}:
\begin{equation}
	\sigma^{NN}_{inel}(s_{NN}) = A_\sigma + B_\sigma \ln^n(s_{NN}),
	\label{eq:sigma_NN}
\end{equation}
where  $A_\sigma = 29.8 \, \text{GeV}^{-2}$, $B_\sigma = 0.038 \, \text{GeV}^{-2}$, $n= 2.43 $. $\Gamma(s_{NN}, \mathbf{b})$ denotes the Fourier transform of the proton-proton elastic scattering amplitude, which is parameterized by an exponential form factor \cite{Frankfurt:2006jp}. The inverse slope $b_0\equiv b_0(\sqrt{s_{NN}})$ exhibits a functional dependence on the $pp$ center-of-mass energy. The experimental data were fitted to the functional form $b_0(\sqrt{s_{NN}}) = A_b + B_b \ln(\sqrt{s_{NN}}) + C_b \ln^2(\sqrt{s_{NN}})$,  yielding  $A_b = 9.81 \, \text{GeV}^{-2}$, $B_b = 0.211 \, \text{GeV}^{-2}$, $C_b= 0.00185 \, \text{GeV}^{-2}$. For $pp$ collisions at the LHC (14 TeV) and FCC (100 TeV), we have $b_0=20.6, 24.5 \, \text{GeV}^{-2}$ \cite{Shao:2022cly}. The remaining the input parameters introduced above are summarized in Sec.~\ref{sec3}.

The conventional photon flux formulation, employed in this analysis and commonly adopted in the literature \cite{Klasen:2008mh,Chen:2025gwp,Shao:2022cly}, is grounded in the EDFF of the emitting hadron. For an ion beam of charge 
$Z$, the photon number density at impact parameter $b$, derived from the EDFF, is given by,

\begin{equation}
	\label{fra:EDFF}  
	N_{\gamma/Z}^{\mathrm{EDFF}} (E_\gamma, b) = \frac{Z^2 \alpha}{\pi^2} \frac{\xi^2}{b^2} \biggl[ K_1^2 (\xi) + \frac{1}{\gamma_{\mathrm{L}}^2} K_0^2 (\xi) \biggr]
\end{equation}
where $\xi =E_\gamma b/\gamma_L$, $K_{i}$ are modified Bessel functions, $\gamma_{\mathrm{L}} = E_{beam} /m_{p,N}$ is the Lorentz factor of the initial-state proton or nucleus.

\begin{figure*}[htbp]
	\centering
	\includegraphics[height=6cm,width=15cm]{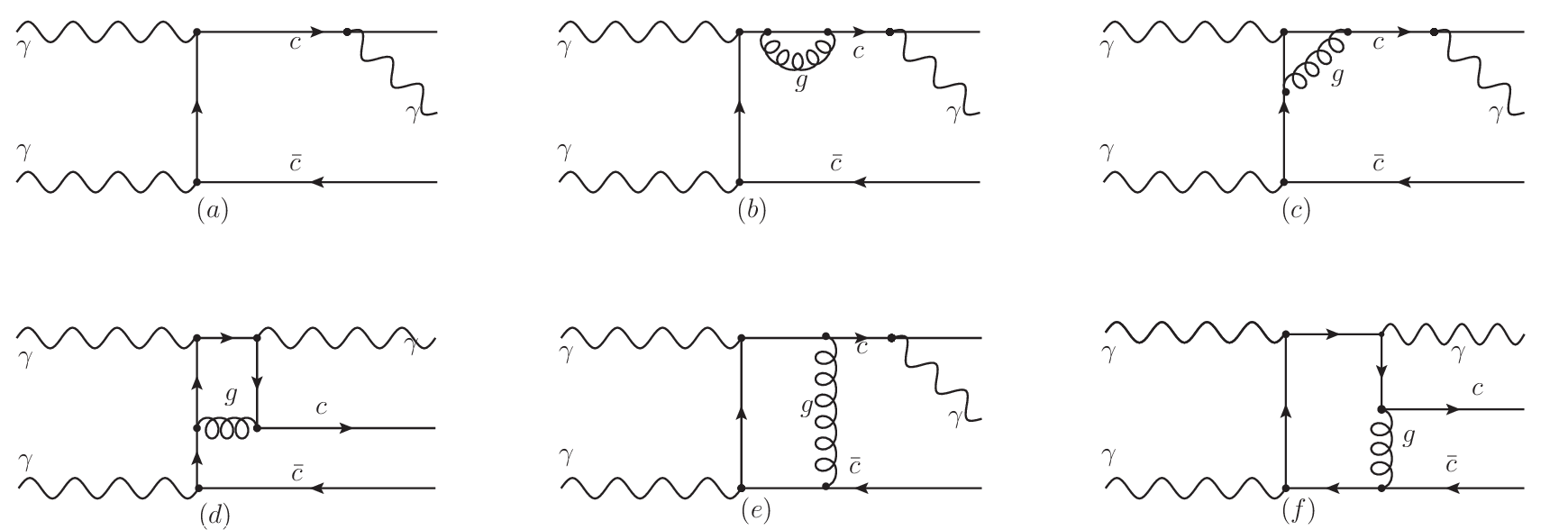}
\caption{
	Diagram (a) denotes a typical tree-level contribution, while diagrams (b)--(f) represent typical one-loop virtual corrections.
}
\label{fig:feynman}
\end{figure*}

Figure~\ref{fig:feynman} displays the representative Feynman diagrams for the exclusive process.
At LO, there are six tree-level diagrams, one representative tree-level topology is displayed in Fig.~\ref{fig:feynman}(a). At NLO in QCD, the virtual corrections consist of 48 one-loop diagrams: 12 self-energy diagrams, 18 triangle diagrams, 12 box diagrams, and six pentagon diagrams. In the adopted scheme presented below, the self-energy diagrams on external quark legs do not need to be calculated. Some typical one-loop topologies are illustrated in Figs.~\ref{fig:feynman}(b)--\ref{fig:feynman}(f).

Owing to the exclusive nature of the process, the QCD NLO contribution arises solely from virtual corrections, leading to a well-defined modification of the hard scattering amplitude. In our NLO calculations, we encountered both ultraviolet (UV) and infrared (IR) divergences, which were regularized via dimensional regularization in $D = 4 - 2\epsilon$ dimensions. The UV divergences were removed through renormalization of the parameters and external fields in the tree-level amplitude. We implemented the on-mass-shell (OS) scheme for the renormalization of the charm quark wave function ($Z_2$) and mass ($Z_m$), as well as the gluon wave function ($Z_3$). For the renormalization of the strong-coupling constant ($Z_g$), we used the modified minimal-subtraction ($\overline{\text{MS}}$) scheme.

The corresponding one-loop renormalization constants are given by \cite{He:2024lrb},
\begin{align}
	\delta Z_2^{\text{OS}} &= -C_F \frac{\alpha_s}{4\pi} N_\epsilon \left( \frac{1}{\epsilon_{\text{UV}}} + \frac{2}{\epsilon_{\text{IR}}} + 4 \right), \\
	\delta Z_m^{\text{OS}} &= -3C_F \frac{\alpha_s}{4\pi} N_\epsilon \left( \frac{1}{\epsilon_{\text{UV}}} + \frac{4}{3} \right), \\
	\delta Z_3^{\text{OS}} &= \frac{\alpha_s}{4\pi} N_\epsilon \left[ (\beta_0' - 2C_A) \left( \frac{1}{\epsilon_{\text{UV}}} - \frac{1}{\epsilon_{\text{IR}}} \right) \right. \\
	&\quad \left. - \frac{4}{3} T_F (n_f - n_l) \frac{1}{\epsilon_{\text{UV}}} \right] , \nonumber\\
	\delta Z_g^{\overline{\text{MS}}} &= -\frac{\beta_0}{2} \frac{\alpha_s}{4\pi} N_\epsilon \left( \frac{1}{\epsilon_{\text{UV}}} + \ln \frac{m_c^2}{\mu_r^2} \right) ,
\end{align}
where $N_{\varepsilon} = (4\pi\mu_r^2/m_c^2)^{\varepsilon}/\Gamma(1-\varepsilon)$. $\mu_r$ is the renormalization scale. 
Here, $\epsilon_{\text{UV}}$ and $\epsilon_{\text{IR}}$ represent the dimensional regularization parameters corresponding to the UV and IR singularities, respectively.
$C_F = (N_c^2-1)/2N_c$, $C_A = N_c$, $T_F = 1/2$ . $n_f$ and $n_l$ are the numbers of active quark flavors and light quark flavors, $n_l=3$ and $n_f=4$.
One loop QCD $\beta$ function $\beta_0 = 11N_c/3 - 2n_f/3$. $\beta_0'$ is scheme-dependent and emerges from $\beta_0$ by replacing $n_f$ with $n_l$.

In the calculation of NLO corrections, we first employ the QGRAF package~\cite{Nogueira:1991ex} to automatically generate all relevant Feynman diagrams. The Dirac and color algebra are subsequently handled through FORM~\cite{Vermaseren:2000nd}. The amplitudes are reduced to a set of master integrals using Reduze~2~\cite{vonManteuffel:2012np} and FIRE6~\cite{Smirnov:2019qkx}. Analytical expressions for the master integrals are obtained with Package-X~2.0~\cite{Patel:2016fam} and QCDloop~\cite{Ellis:2007qk}, and the phase-space Monte Carlo integrations are performed using the Cuba library~\cite{Hahn:2004fe,Hahn:2014fua}.

\section{Numerical Results and Discussion}\label{sec3}
 
\subsection{Input parameters}\label{sec3.1}

Our numerical analysis considers exclusive $J/\psi+\gamma$ production in UPCs at the LHC and FCC. The collision systems, energies and the corresponding $\sigma^{NN}_{\text{inel}}$ (calculated using Eq.~\eqref{eq:sigma_NN}) are listed in Table~\ref{tab:sigma_NN}.

\begin{table}[htbp]
	\centering
	\caption{Nucleon--nucleon inelastic cross sections $\sigma^{NN}_{\text{inel}}$ at different collision energies.}
	\label{tab:sigma_NN}
	\begin{tabular}{lcc}
		\toprule
		System & $\sqrt{s_{NN}}$ (TeV) & $\sigma^{NN}_{\text{inel}}$ (mb) \\
		\midrule
		\multicolumn{3}{c}{HL-LHC} \\
		\midrule
		Pb--Pb & 5.52  & 68.50 \\
		Xe--Xe & 5.86  & 68.83 \\
		Kr--Kr & 6.46  & 69.90 \\
		Ar--Ar & 6.30  & 69.63 \\
		Ca--Ca & 7.00  & 70.80 \\
		O--O   & 7.00  & 70.80 \\
		$p$--Pb  & 8.80  & 73.30 \\
		$p$--$p$ & 14.00 & 79.04 \\
		\midrule
		\multicolumn{3}{c}{FCC} \\
		\midrule
		Pb--Pb & 39.4  & 93.03 \\
		$p$--Pb  & 62.8  & 100.02 \\
		$p$--$p$ & 100.0 & 107.42 \\
		\bottomrule
	\end{tabular}
\end{table}

In our numerical simulations, the initial geometry of colliding nuclei is defined by the Woods–Saxon density profile given in Eq.~\eqref{eq:fermi3}. The nuclear parameters $A$, $Z$, and the Woods--Saxon parameters $R_A$, $a_A$, and $w_A$ are taken from Refs.~\cite{Loizides:2017ack,DeJager:1974liz,DeVries:1987atn}, 
with $R_A = 1.2\,\mathrm{fm}\,A^{1/3}$~\cite{Baur:2001jj}. 
The corresponding values are listed in Table~\ref{tab:ws1}.

\begin{table}[t]
	\centering
	\caption{Parameters of the Woods--Saxon nuclear density profile and beam settings used in the numerical calculations~\cite{Loizides:2017ack,DeJager:1974liz,DeVries:1987atn,Baur:2001jj}.}
	\label{tab:ws1}
	\begin{tabular}{lcccccccc}
		\toprule
		Nucleus & $E_{\rm beam}$ (TeV) & $\gamma_L = E_{\rm beam}/m_{p,N}$ & $R_A$ (fm) & $a_A$ (fm)   & $w_A$ & $A$ & $Z$ \\
		\midrule
		p   & 7    & 7460  & 0.7 & --           & --     & --  & -- \\
		p   & 50   & 53287 & 0.7 & --         & --     & --  & -- \\
		Pb  & 2.76 & 2962  & 7.1 & 0.549    & 0      & 208 & 82 \\
		Pb  & 19.7  & 21148 & 7.1 & 0.549      & 0      & 208 & 82 \\
		Xe  & 2.93  & 3145  & 6.1 & 0.590    & 0      & 129 & 54 \\
		Kr  & 3.23 & 3467  & 5.1 & 0.500   & 0      & 78  & 36 \\
		Ar  & 3.15 & 3381  & 4.1 & 0.586 & $-0.161$ & 40  & 18 \\
		Ca  & 3.5  & 3757  & 4.1 & 0.00 & $-0.161$ & 40  & 20 \\
		O   & 3.5  & 3757  & 3.1 & 0.00   & $-0.051$ & 16  & 8 \\
		\bottomrule
	\end{tabular}
	\vspace{2mm}
	\raggedright
	\\
    \footnotesize
	Input nucleon masses: $m_p = 0.9383~\mathrm{GeV}$, $m_N = 0.9315~\mathrm{GeV}$.
\end{table}

The input parameters for alternative heavy-ion collisions at varying nucleon-nucleon center-of-mass energies are taken directly from the gamma-UPC framework \cite{Shao:2022cly}. Furthermore, we incorporate the parameters for the NLO calculations of heavy quarkonium production through photon-photon interactions,

\begin{equation}
	\alpha = 1/137.065,\; m_c = 1.5 \, \text{GeV}, \;|R_{S}(0)|^2 = 0.81 \, \text{GeV}^{3} 
\end{equation}
The value of the radial wave function at the origin is taken from the B-T potential model\cite{Eichten:1995ch}, which has been widely adopted in both LO and NLO studies of heavy quarkonium production\cite{Liu:2003jj,He:2024lrb,Campbell:2007ws,Gong:2008sn,Shao:2014yta,Chao:2012iv,Ma:2010jj}.

The running two-loop  \(\alpha_s\) as a function of $\mu_r$ reads

\begin{equation}
	\alpha_s(\mu_r) = \frac{4\pi}{\beta_0 \ln \mu_r^2 / \Lambda_{\text{QCD}}^2} \left[ 1 - \frac{\beta_1 \ln \ln \mu_r^2 / \Lambda_{\text{QCD}}^2}{\beta_0^2 \ln \mu_r^2 / \Lambda_{\text{QCD}}^2} \right]
\end{equation}
where the two-loop QCD $\beta$ function  
$\beta_1 = \frac{34}{3} C_A^2 - 4 C_F T_{F} n_f - \frac{20}{3} C_A T_{F} n_f$. We take $n_f = 4$,  
$\Lambda_{\text{QCD}} = 326 \, \text{MeV}$~\cite{Pumplin:2002vw}.

\subsection{$pp$, $p$--Pb, and Pb--Pb collisions at HL-LHC and FCC}\label{sec3.2}

Table~\ref{tab:LHC} and Table~\ref{tab:FCC} present the total cross sections for exclusive 
\(J/\psi+\gamma\) production in ultraperipheral $pp$, $p$–Pb, and Pb–Pb collisions 
at the LHC and the FCC. Results are given at both LO and NLO accuracy together with 
the corresponding \(K\) factors, defined as
$K=\sigma_{\mathrm{NLO}}/\sigma_{\mathrm{LO}}$. To quantify the perturbative uncertainty from uncalculated higher-order contributions, 
two types of renormalization-scale choices are adopted. One is a dynamical scale linked to the photon–photon invariant mass, \(\mu_r\sim\sqrt{\hat s}=\sqrt{x_1x_2s}\). The other is based on the transverse kinematics, \(\mu_r\sim m_T=\sqrt{4m_c^2+p_T^2}\). In each case, the scale is varied by a factor of two around its central value. The predicted cross sections exhibit a 
pronounced dependence on the transverse-momentum cut imposed on the final-state 
\(J/\psi\). To illustrate this effect, we consider two representative thresholds, $p_T>0~\mathrm{GeV}$ and $p_T>2~\mathrm{GeV}$.

\begin{table}[!htbp]
	\caption{Total cross sections for exclusive $J/\psi + \gamma$ production via photon–photon fusion in UPCs at LO and NLO for different collision systems at LHC($\sqrt{s}$ is in units of TeV). The cross sections are given in units of fb ($pp$), pb ($p$–Pb), and nb (Pb–Pb), respectively, under two transverse-momentum cuts: $p_T > 0$ GeV and $p_T > 2$ GeV. The dependence on the renormalization scale $\mu_r$ is shown for six different choices.}
    \label{tab:LHC}
	\centering
	\small
	\renewcommand{\arraystretch}{1.0} 
	\begin{tabular}{@{}lccrrrrrr@{}}
		\toprule
		\multirow{2}{*}{\shortstack{Collision\\ system \\ ($\sqrt{s}$)}} 
		& \multirow{2}{*}{\centering\arraybackslash \shortstack{$p_T$ cut \\ (GeV)}} 
		& \multirow{2}{*}{Order} 
		& \multicolumn{6}{c}{$\mu_r$} \\
		\cmidrule(lr){4-9}
		& & 
		& $\sqrt{\hat s}$ 
		& $\frac{1}{2}\sqrt{\hat s}$ 
		& $2\sqrt{\hat s}$ 
		& $m_T$ 
		& $\frac{1}{2}m_T$ 
		& $2m_T$ \\
		\midrule
		\multirow{6}{*}{\begin{tabular}{@{}l@{}}$pp$ \\ ($14$~TeV)\end{tabular}}
		& \multirow{3}{*}{$>0$} & LO & \multicolumn{6}{c}{155.06} \\
		& & NLO & 124.44 & 114.83 & 130.09 & 120.62 & 107.49 & 127.68 \\
		& & $K$ & 0.80 & 0.74 & 0.84 & 0.78 & 0.69 & 0.82 \\
		\cmidrule(lr){2-9}
		& \multirow{3}{*}{$>2$} & LO & \multicolumn{6}{c}{14.61} \\
		& & NLO & 9.40 & 8.10 & 10.25 & 8.40 & 6.31 & 9.56 \\
		& & $K$ & 0.64 & 0.55 & 0.70 & 0.57 & 0.43 & 0.65 \\
		\midrule
		\multirow{6}{*}{\begin{tabular}{@{}l@{}}$p$--Pb \\ ($8.8$~TeV)\end{tabular}}
		& \multirow{3}{*}{$>0$} & LO & \multicolumn{6}{c}{448.50} \\
		& & NLO &361.00&333.14&378.69&350.18&316.46& 373.12 \\
		& & $K$ & 0.80 & 0.74 & 0.84 & 0.78 & 0.70 & 0.83 \\
		\cmidrule(lr){2-9}
		& \multirow{3}{*}{$>2$} & LO & \multicolumn{6}{c}{38.43} \\
		& & NLO & 24.25 & 21.37 & 27.05 & 21.37 & 15.83 & 25.94 \\
		& & $K$ & 0.63 & 0.56 & 0.70 & 0.56 & 0.41 & 0.67 \\
		\midrule
		\multirow{6}{*}{\begin{tabular}{@{}l@{}}Pb--Pb \\ ($5.52$~TeV)\end{tabular}}
		& \multirow{3}{*}{$>0$} & LO & \multicolumn{6}{c}{1012.03} \\
		& & NLO & 815.88 & 753.97 & 850.61 & 795.64 & 714.97 & 843.96 \\
		& & $K$ & 0.81 & 0.75 & 0.84 & 0.79 & 0.71 & 0.83 \\
    	\cmidrule(lr){2-9}
		& \multirow{3}{*}{$>2$} & LO & \multicolumn{6}{c}{70.33} \\
		& & NLO & 43.67 & 37.00 & 48.07 & 38.39 & 28.10 & 44.63 \\
		& & $K$ & 0.62 & 0.53 & 0.68 & 0.55 & 0.40 & 0.63 \\
		\bottomrule
	\end{tabular}
\end{table}

\begin{table}[!htbp]
	\centering
	\caption{Total cross sections for exclusive $J/\psi + \gamma$ production via photon–photon fusion in UPCs at LO and NLO. Results are presented for $pp$, $p$--Pb, and Pb--Pb collision systems at FCC($\sqrt{s}$ is in units of TeV). The cross sections are given in units of fb ($pp$), pb ($p$--Pb), and nb (Pb--Pb), under two transverse-momentum cuts: $p_T > 0\,\mathrm{GeV}$ and $p_T > 2\,\mathrm{GeV}$. The dependence on the renormalization scale $\mu_r$ is shown for six different choices.}
    \label{tab:FCC}
	\small
	\renewcommand{\arraystretch}{1.0} 
	\begin{tabular}{@{}lccrrrrrr@{}}
		\toprule
		\multirow{2}{*}{\shortstack{Collision\\ system \\ ($\sqrt{s}$)}} 
		& \multirow{2}{*}{\centering\arraybackslash \shortstack{$p_T$ cut \\ (GeV)}} 
		& \multirow{2}{*}{Order} 
		& \multicolumn{6}{c}{$\mu_r$} \\
		\cmidrule(lr){4-9}
		& & 
		& $\sqrt{\hat s}$ 
		& $\frac{1}{2}\sqrt{\hat s}$ 
		& $2\sqrt{\hat s}$ 
		& $\frac{1}{2}m_T$ 
		& $m_T$ 
		& $2m_T$ \\
		\midrule
		\multirow{6}{*}{\begin{tabular}{@{}l@{}}$pp$ \\ ($100$~TeV)\end{tabular}}
		& \multirow{3}{*}{$>0$} & LO &\multicolumn{6}{c}{289.24} \\
		& & NLO & 232.88 & 215.23 & 243.17 & 202.04 & 225.86 & 238.70 \\
		& & $K$ & 0.80 & 0.74 & 0.84 & 0.70 & 0.78 & 0.82 \\
		\cmidrule(lr){2-9}
		& \multirow{3}{*}{$>2$} & LO&\multicolumn{6}{c}{29.10}  \\
		& & NLO & 18.55 & 15.92 & 20.25 & 12.29 & 16.47 & 18.87 \\
		& & $K$ & 0.64 & 0.55 & 0.70 & 0.42 & 0.57 & 0.65 \\
		\midrule
		\multirow{6}{*}{\begin{tabular}{@{}l@{}}$p$--Pb \\ ($62.8$~TeV)\end{tabular}}
		& \multirow{3}{*}{$>0$} & LO &\multicolumn{6}{c}{1081.44}  \\
		& & NLO & 869.74 & 798.83 & 907.39 & 759.00 & 843.65 & 889.39 \\
		& & $K$ & 0.80 & 0.74 & 0.84 & 0.70 & 0.78 & 0.82 \\
		\cmidrule(lr){2-9}
		& \multirow{3}{*}{$>2$} & LO &\multicolumn{6}{c}{102.61}  \\
		& & NLO & 66.28 & 55.22 & 70.71 & 43.35 & 57.63 & 66.05 \\
		& & $K$ & 0.65 & 0.54 & 0.69 & 0.42 & 0.56 & 0.64 \\
		\midrule
		\multirow{6}{*}{\begin{tabular}{@{}l@{}}Pb--Pb \\ ($39.4$~TeV)\end{tabular}}
		& \multirow{3}{*}{$>0$} & LO &\multicolumn{6}{c}{3886.12} \\
		& & NLO & 3112.05 & 2867.82 & 3277.52 & 2687.11 & 3034.00 & 3196.24 \\
		& & $K$ & 0.80 & 0.74 & 0.84 & 0.69 & 0.78 & 0.82 \\
		\cmidrule(lr){2-9}
		& \multirow{3}{*}{$>2$} & LO &\multicolumn{6}{c}{340.83}  \\
		& & NLO & 214.73 & 180.83 & 233.99 & 141.11 & 189.13 & 217.12 \\
		& & $K$ & 0.63 & 0.53 & 0.69 & 0.41 & 0.55 & 0.64 \\
		\bottomrule
	\end{tabular}
\end{table}

At LO, the total cross sections are independent of the renormalization scale
$\mu_r$, as expected for a purely QED-induced process.
More importantly, their magnitudes exhibit a clear hierarchy among different
collision systems,
\begin{equation}
	\sigma_{\mathrm{PbPb}} \gg \sigma_{p\text{Pb}} \gg \sigma_{pp},
\end{equation}
which reflects the strong enhancement of the equivalent photon flux associated with the nuclear charge.
Accordingly, the cross sections are of the order of nb in Pb--Pb collisions, pb in $p$--Pb collisions, and fb in $pp$ collisions. This hierarchy remains unchanged after including NLO QCD corrections.

As a consistency check, we compare the relative sizes of the cross sections
among different collision systems with those reported in the literature for
other photon–photon–initiated quarkonium production processes.
For instance, Ref.~\cite{Goncalves:2018yxc} investigate the $\eta_c$ production by photon-photon interactions in $pp$ and $p$--Pb collisions at the LHC energies. Ref.~\cite{Shao:2022cly} presents results for
$\gamma\gamma \to J/\psi J/\psi$ production in $pp$, $p$--Pb, and Pb--Pb collisions.
Despite the different final states, the ratios among Pb--Pb, $p$--Pb, and $pp$
cross sections exhibit a similar hierarchical pattern, originating from the
same nuclear-charge $Z$ enhancement of the equivalent photon flux in UPCs. This agreement supports the reliability of our numerical results.

We find that NLO QCD corrections have a sizable impact on the total cross
sections for all collision systems considered.
For the scale choice $\mu_r = \sqrt{x_1 x_2 s}$ and without imposing a
transverse-momentum cut ($p_T > 0$~GeV), the $pp$ cross section decreases from
155.06~fb at LO to 124.44~fb at NLO, corresponding to a $K$ factor of 0.80.
Similarly, the LO cross sections in $p$--Pb and Pb--Pb collisions are reduced by about $20\%$ once NLO corrections are included, yielding $K$ factors close to 0.8.
This indicates that the negative NLO contributions are sizable and exhibit a consistent behavior across different collision systems. For $p_T>2$~GeV and the scale choice $\mu_r = \sqrt{4m_c^2+p_T^2}$, the NLO QCD corrections become more pronounced, reducing the LO cross sections by several
tens of percent in both $pp$ and Pb--Pb collisions. In Ref.~\cite{Chen:2025gwp}, the $\mathcal{O}(\alpha_s)$ corrections reported in
Table~II are also negative and of comparable relative size, lowering $\sigma_{\mathrm{LO}}$ by about $50\%$ in both collision systems.
Moreover, Ref.~\cite{Goncalves:2023sts} presents Pb--Pb cross sections at LHC
energies for the $J/\psi+\gamma$ final state. After taking into account the uncertainties associated with the LDMEs, our LO predictions are found to be consistent with their results.

The residual renormalization-scale dependence of the NLO results provides an
estimate of the perturbative uncertainty.
For $p_T>0$~GeV, the NLO cross section in $pp$ collisions at  HL-LHC varies from 107.49 to
130.09~fb under typical scale variations, corresponding to a relative
uncertainty of about $+8\%$ to $-11\%$ with respect to the central value
120.62~fb.
When a more stringent cut $p_T>2$~GeV is imposed, the cross section spans
6.31–10.25~fb and the associated uncertainty increases to
$+22\%$ to $-25\%$.
A similar pattern is observed in $p$--Pb and Pb--Pb collisions.

The $K$ factor exhibits a corresponding scale dependence.
For $p_T>0$~GeV, it ranges between 0.69 and 0.82, while in the $p_T>2$~GeV region
it decreases to 0.43–0.65. We find that the total cross section decreases when a transverse-momentum cut is imposed. More importantly, the relative size of the NLO correction is enhanced in the presence of a $p_T$ cut, indicating an increased sensitivity of the observable to higher-order QCD radiation under kinematic constraints.

\begin{figure*}[htbp]
	\centering
	\includegraphics[height=11cm,width=17cm]{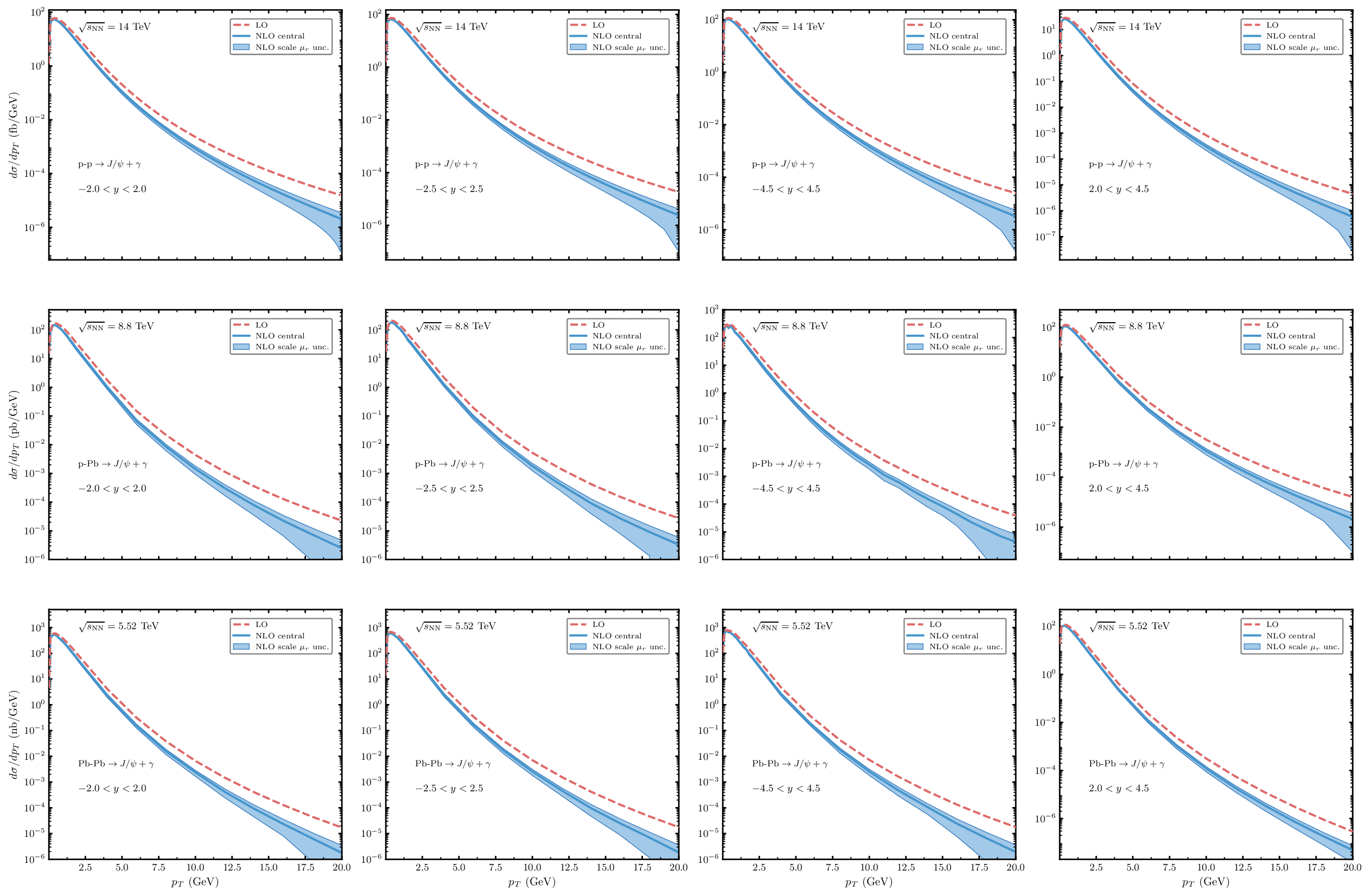}
	\caption{Differential cross sections  $\mathrm{d}\sigma / \mathrm{d}p_T$ in different rapidity ranges at HL-LHC. The dashed line represents the LO results. The shaded band corresponds to the NLO results, and the band is obtained by varying the scale between $\mu_r/2$ and $2\mu_r$. The solid line in the band indicates the central value with the renormalization scale set to  $\mu_r = \sqrt{x_1x_2s}$. }
	\label{LHC-pt}
\end{figure*}

\begin{figure*}[htbp]
	\centering
	\includegraphics[height=12cm,width=17cm]{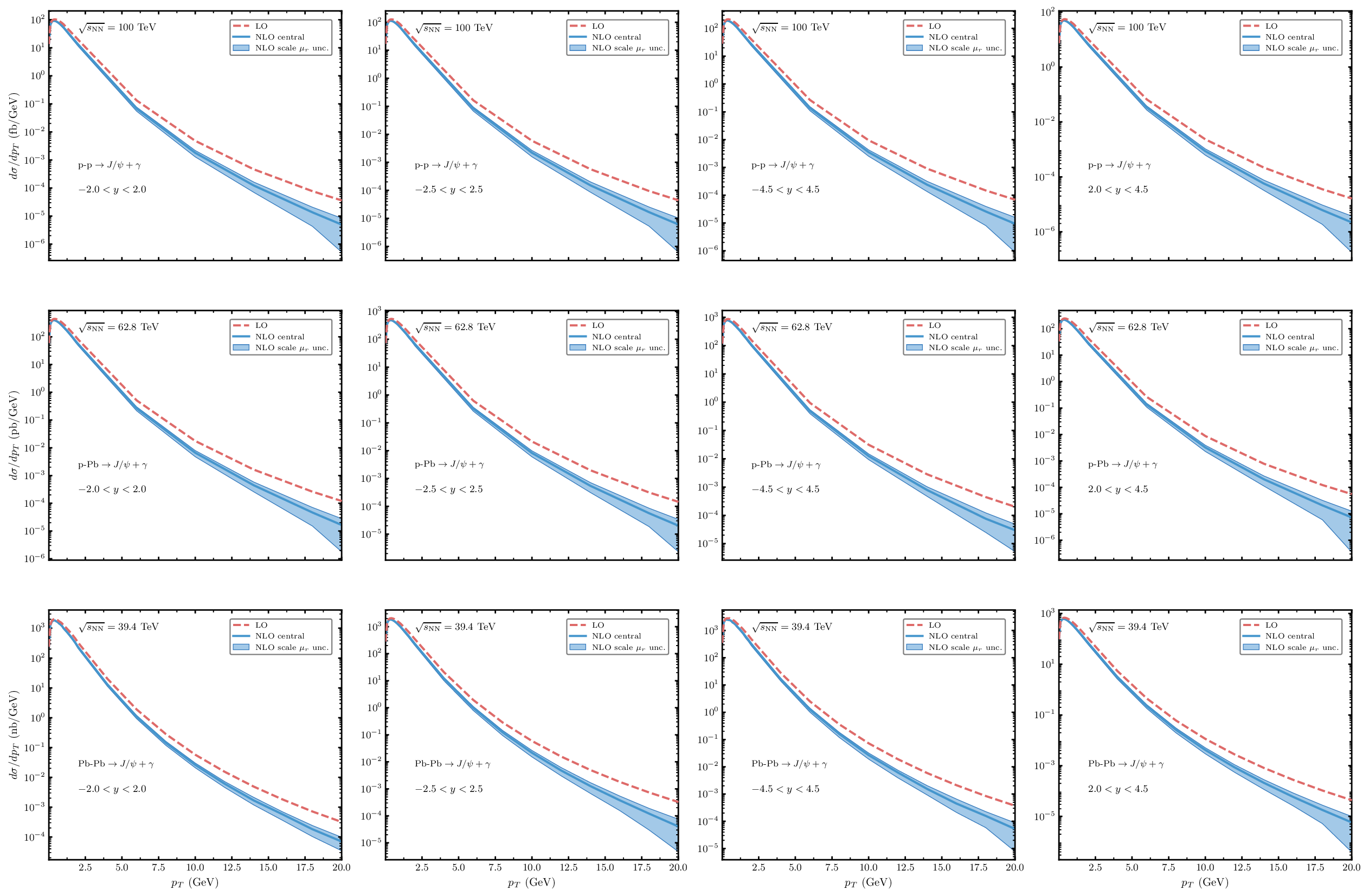}
	\caption{Differential cross sections  $\mathrm{d}\sigma / \mathrm{d}p_T$ in different rapidity ranges at FCC. The dashed line represents the LO results. The shaded band corresponds to the NLO results, and the band is obtained by varying the scale between $\mu_r/2$ and $2\mu_r$. The solid line in the band indicates the central value with the renormalization scale set to  $\mu_r = \sqrt{x_1x_2s}$. }
	\label{FCC-pt}
\end{figure*}

\begin{figure}[htbp]
	\centering
	\includegraphics[height=12cm,width=15cm]{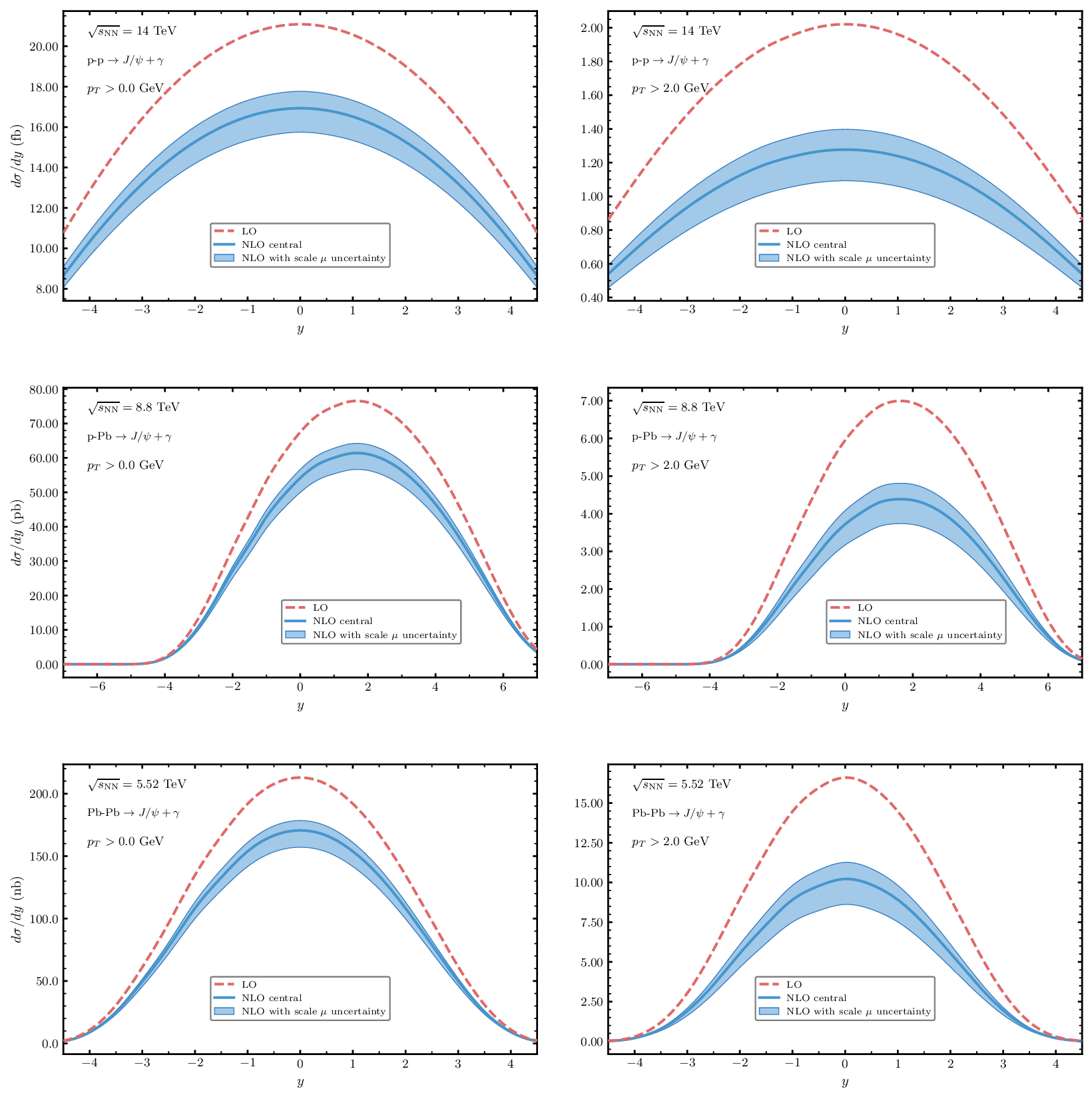}
	\caption{Differential cross sections  $\mathrm{d}\sigma / \mathrm{d}y$ at HL-LHC. The dashed line represents the LO results. The shaded band corresponds to the NLO results, and the band is obtained by varying the scale between $\mu_r/2$ and $2\mu_r$. The solid line in the band indicates the central value with the renormalization scale set to  $\mu_r = \sqrt{x_1x_2s}$.}
	\label{LHC-y}
\end{figure}

\begin{figure}[htbp]
	\centering
	\includegraphics[height=12cm,width=15cm]{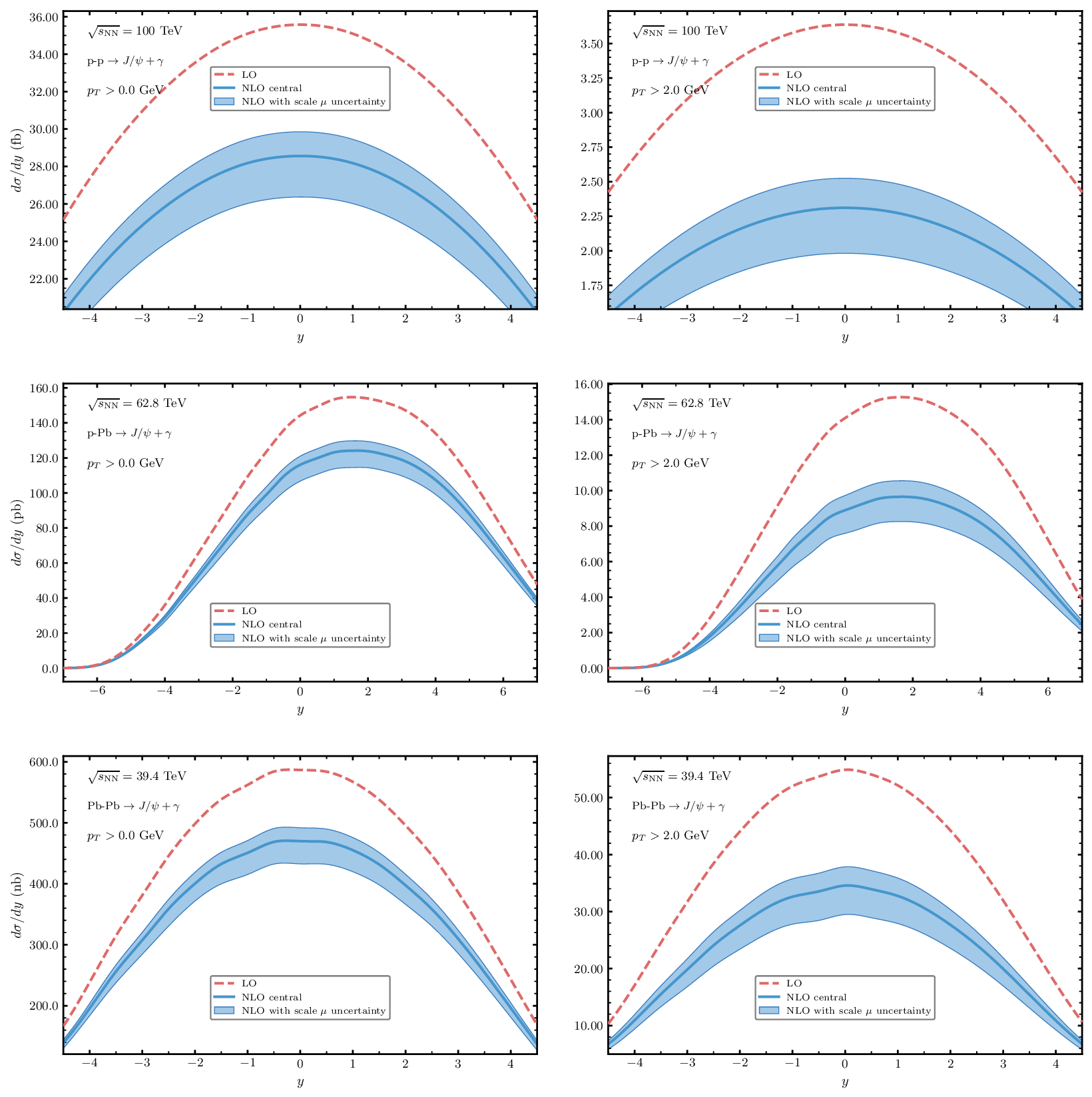}
	\caption{Differential cross sections  $\mathrm{d}\sigma / \mathrm{d}y$ at FCC. The dashed line represents the LO results. The shaded band corresponds to the NLO results, and the band is obtained by varying the scale between $\mu_r/2$ and $2\mu_r$. The solid line in the band indicates the central value with the renormalization scale set to  $\mu_r = \sqrt{x_1x_2s}$.}
	\label{FCC-y}
\end{figure}

In Figs.~\ref{LHC-pt}--\ref{FCC-y}, we present the $p_T$ and rapidity (denoted by $y$) distributions of $J/\psi$, respectively. 
The $p_T$ ranges from $0$ GeV to $20$ GeV, and $y$ ranges from $-4.5$ to $4.5$ for $pp$ and Pb--Pb collisions and $-7.0$ to $7.0$ for $p$--Pb collisions. The NLO theoretical uncertainties are estimated by varying the scale between $\mu_r/2$ and $2\mu_r$. While the NLO corrections maintain the overall shape of the distributions, they result in a significant suppression that intensifies with increasing $p_T$, with the low-$p_T$ region remaining largely unaffected.

As seen in Figs. \ref{LHC-pt} and \ref{FCC-pt}, most contributions are localized at the low-$p_T$ regions. For the rapidity distribution, a particularly pronounced suppression is observed in the central region ($|y|<2.5$). The spectra are symmetric in $pp$ and Pb--Pb collisions, reflecting the identical photon flux configurations from the two projectiles. In contrast, the $p$--Pb system exhibits a pronouncedly asymmetric rapidity profile, reflecting the imbalance of photon fluxes between the proton and the nucleus. One contributing factor is the boost from the nucleon--nucleus center-of-mass frame to the laboratory frame. As discussed in Ref.~\cite{Citron:2018lsq}, the resulting shift of the central rapidity can be approximated as
\begin{equation}
 	y_{NN} \approx \frac{1}{2}\log\!\left(\frac{Z_1 A_2}{A_1 Z_2}\right) \approx
 	\begin{cases}
 		0, & \text{$pp$, Pb--Pb}, \\
 		0.465, & \text{$p$--Pb}.
 	\end{cases}
\end{equation}
In addition, the photon flux from the nucleus is strongly enhanced by the $Z^2$ factor, in agreement with the rapidity peak near $y=1.5$ found in Ref.~\cite{Goncalves:2018yxc}.
Similar asymmetric features have been observed in previous studies of photon-induced quarkonium production in asymmetric collision systems such as $p$--Pb~\cite{ALICE:2014eof}, where the rapidity distributions are correspondingly shifted toward the Pb-emission direction.

It should be mentioned that, in a realistic hadronic collision environment, the same
exclusive final state may also receive contributions from diffractive background
processes, such as pomeron-pomeron induced production. The associated production of
$J/\psi+\gamma$ through double-pomeron exchange has been investigated in
Refs.~\cite{Xu:1998rp,GayDucati:2009rr}, and related quarkonium-plus-photon production
in coherent hadron-hadron interactions was discussed in Ref.~\cite{Goncalves:2012bt}.
Such contributions are not included in the present calculation, since they correspond
to a different production mechanism from the photon-induced process considered here.
Nevertheless, these backgrounds may have different transverse-momentum and diffractive
characteristics compared with the ultraperipheral photon-induced signal. Therefore,
they may be reduced experimentally by applying suitable cuts on the transverse
momenta of the final-state particles~\cite{Lebiedowicz:2023mhe}.

\subsection{ $p$--A, and A--A collisions at HL-LHC}\label{sec3.3}

In the previous Subsection we discussed the $p$--Pb case, where the production rate is driven by the photon flux of a single proton. We next consider $p$--A and A--A UPCs at the HL-LHC, where the coherent electromagnetic fields of nuclei substantially enhance the photon luminosity. In these systems the cross section scales approximately with the nuclear charge as $Z^2$ for $p$--A and $Z^4$ for A--A collisions, making UPCs with medium and heavy nuclei particularly interesting despite their lower hadronic luminosities.

The systematics observed in Tables~\ref{tab:proton–nucleus} and 
\ref{tab:nucleus–nucleus} further confirm that the production rate is 
primarily governed by the nuclear charge dependence of the photon flux. 
Since the equivalent photon spectrum scales approximately as $Z^2$, the 
total cross section in nucleus–nucleus UPCs follows a $Z^4$ enhancement 
pattern, while $p$--A systems scale as $Z^2$.

\begin{table}[!htbp]
	\centering
	\caption{Total cross sections for exclusive $J/\psi + \gamma$ production via photon–photon fusion in UPCs at LO and NLO for proton–nucleus collision systems ($\sqrt{s}$ is in units of TeV). The cross sections are given in units of pb, under two transverse-momentum cuts: $p_T > 0$ GeV and $p_T > 2$ GeV. The dependence on the renormalization scale $\mu_r$ is shown for six different choices.}
	\label{tab:proton–nucleus}
	\small
	\renewcommand{\arraystretch}{0.75}    
	\setlength{\tabcolsep}{4pt}           
	\begin{tabular}{@{}lccrrrrrr@{}}
		\toprule
		\multirow{2}{*}{\shortstack{Collision\\ system \\ ($\sqrt{s}$)}} 
		& \multirow{2}{*}{\centering\arraybackslash \shortstack{$p_T$ cut \\ (GeV)}} 
		& \multirow{2}{*}{Order} 
		& \multicolumn{6}{c}{$\mu_r$} \\
		\cmidrule(lr){4-9}
		& & 
		& $\sqrt{\hat s}$ 
		& $\frac{1}{2}\sqrt{\hat s}$ 
		& $2\sqrt{\hat s}$ 
		& $\frac{1}{2}m_T$ 
		& $m_T$ 
		& $2m_T$ \\
		\midrule
		\multirow{6}{*}{\begin{tabular}{@{}l@{}}p--Xe \\ ($9.06$~TeV)\end{tabular}} & $>0$ & LO & \multicolumn{6}{c}{207.83} \\
		&      & NLO & 168.36 & 154.81 & 174.65 & 144.49 & 163.48 & 171.54 \\
		&      & $K$ & 0.81 & 0.74 & 0.84 & 0.70 & 0.79 & 0.83 \\
		& $>2$ & LO & \multicolumn{6}{c}{17.83} \\
		&      & NLO & 11.30 & 9.52 & 12.23 & 7.33 & 10.06 & 11.40 \\
		&      & $K$ & 0.63 & 0.53 & 0.69 & 0.41 & 0.56 & 0.64 \\
		\midrule
		\multirow{6}{*}{\begin{tabular}{@{}l@{}}p--Kr \\ ($9.51$~TeV)\end{tabular}}& $>0$ & LO & \multicolumn{6}{c}{99.25} \\
		&      & NLO & 79.73 & 74.07 & 83.04 & 68.91 & 77.09 & 80.53 \\
		&      & $K$ & 0.80 & 0.75 & 0.84 & 0.69 & 0.78 & 0.81 \\
		& $>2$ & LO & \multicolumn{6}{c}{8.75} \\
		&      & NLO & 5.51 & 4.69 & 6.01 & 3.62 & 4.93 & 5.66 \\
		&      & $K$ & 0.63 & 0.54 & 0.69 & 0.41 & 0.56 & 0.65 \\
		\midrule
		\multirow{6}{*}{\begin{tabular}{@{}l@{}}p--Ar \\ ($9.39$~TeV)\end{tabular}}& $>0$ & LO & \multicolumn{6}{c}{26.58} \\
		&      & NLO & 21.52 & 19.78 & 22.36 & 18.48 & 20.82 & 22.00 \\
		&      & $K$ & 0.81 & 0.74 & 0.84 & 0.70 & 0.78 & 0.83 \\
		& $>2$ & LO & \multicolumn{6}{c}{2.31} \\
		&      & NLO & 1.46 & 1.23 & 1.60 & 0.94 & 1.29 & 1.47 \\
		&      & $K$ & 0.63 & 0.53 & 0.69 & 0.41 & 0.56 & 0.64 \\
		\midrule
		\multirow{6}{*}{\begin{tabular}{@{}l@{}}p--Ca \\ ($9.9$~TeV)\end{tabular}}& $>0$ & LO & \multicolumn{6}{c}{33.82} \\
		&      & NLO & 27.28 & 25.28 & 28.58 & 23.59 & 26.49 & 27.89 \\
		&      & $K$ & 0.81 & 0.75 & 0.85 & 0.70 & 0.78 & 0.82 \\
		& $>2$ & LO & \multicolumn{6}{c}{3.01} \\
		&      & NLO & 1.90 & 1.64 & 2.07 & 1.25 & 1.68 & 1.93 \\
		&      & $K$ & 0.63 & 0.55 & 0.69 & 0.42 & 0.56 & 0.64 \\
		\midrule
		\multirow{6}{*}{\begin{tabular}{@{}l@{}}p--O \\ ($9.9$~TeV)\end{tabular}} & $>0$ & LO & \multicolumn{6}{c}{5.89} \\
		&      & NLO & 4.77 & 4.41 & 4.95 & 4.11 & 4.62 & 4.87 \\
		&      & $K$ & 0.81 & 0.75 & 0.84 & 0.70 & 0.78 & 0.83 \\
		& $>2$ & LO & \multicolumn{6}{c}{0.51} \\
		&      & NLO & 0.32 & 0.27 & 0.35 & 0.20 & 0.28 & 0.33 \\
		&      & $K$ & 0.63 & 0.53 & 0.69 & 0.39 & 0.55 & 0.64 \\
		\bottomrule
	\end{tabular}
\end{table}

\begin{table}[!htbp]
	\centering
	\caption{Total cross sections for exclusive $J/\psi + \gamma$ production via photon–photon fusion in UPCs at LO and NLO for nucleus–nucleus collision systems($\sqrt{s}$ is in units of TeV). The cross sections are given in units of nb (Xe--Xe, Kr--Kr) and pb (Ar--Ar, Ca--Ca, O--O), under two transverse-momentum cuts: $p_T > 0$ GeV and $p_T > 2$ GeV. The dependence on the renormalization scale $\mu_r$ is shown for six different choices.}
	\label{tab:nucleus–nucleus}
	\small
	\renewcommand{\arraystretch}{0.7}    
	\setlength{\tabcolsep}{4pt}           
	\begin{tabular}{@{}lccrrrrrr@{}}
		\toprule
		\multirow{2}{*}{\shortstack{Collision\\ system \\ ($\sqrt{s}$)}} 
		& \multirow{2}{*}{\centering\arraybackslash \shortstack{$p_T$ cut \\ (GeV)}} 
		& \multirow{2}{*}{Order} 
		& \multicolumn{6}{c}{$\mu_r$} \\
		\cmidrule(lr){4-9}
		& & 
		& $\sqrt{\hat s}$ 
		& $\frac{1}{2}\sqrt{\hat s}$ 
		& $2\sqrt{\hat s}$ 
		& $\frac{1}{2}m_T$ 
		& $m_T$ 
		& $2m_T$ \\
		\midrule
		\multirow{6}{*}{\begin{tabular}{@{}l@{}}Xe--Xe \\ ($5.86$~TeV)\end{tabular}} & \multirow{3}{*}{$>0$} & LO &\multicolumn{6}{c}{249.02}  \\
		& & NLO & 202.26 & 185.07 & 210.23 & 174.46 & 195.34 & 206.39 \\
		& & $K$ & 0.81 & 0.74 & 0.84 & 0.70 & 0.78 & 0.83 \\
		\cmidrule(lr){2-9}
		& \multirow{3}{*}{$>2$} & LO  &\multicolumn{6}{c}{18.35}  \\
		& & NLO & 11.57 & 9.76 & 12.64 & 7.38 & 10.14 & 11.78 \\
		& & $K$ & 0.63 & 0.53 & 0.69 & 0.40 & 0.55 & 0.64 \\
		\midrule
		\multirow{6}{*}{\begin{tabular}{@{}l@{}}Kr--Kr \\ ($6.46$~TeV)\end{tabular}} & 
		\multirow{3}{*}{$>0$} & LO &\multicolumn{6}{c}{62.39}  \\
		& & NLO & 50.29 & 46.49 & 52.60 & 44.30 & 48.97 & 51.63 \\
		& & $K$ & 0.81 & 0.74 & 0.84 & 0.71 & 0.78 & 0.83 \\
		\cmidrule(lr){2-9}
		& \multirow{3}{*}{$>2$} & LO &\multicolumn{6}{c}{4.785}  \\
		& & NLO & 3.042 & 2.575 & 3.342 & 1.998 & 2.705 & 3.081 \\
		& & $K$ & 0.64 & 0.54 & 0.70 & 0.42 & 0.57 & 0.64 \\
		\midrule
		\multirow{6}{*}{\begin{tabular}{@{}l@{}}Ar--Ar \\ ($6.3$~TeV)\end{tabular}} & \multirow{3}{*}{$>0$} & LO &\multicolumn{6}{c}{4561.38}   \\
		& & NLO & 3685.27 & 3394.70 & 3846.66 & 3207.65 & 3582.01 & 3775.78 \\
		& & $K$ & 0.81 & 0.74 & 0.84 & 0.70 & 0.79 & 0.83 \\
		\cmidrule(lr){2-9}
		& \multirow{3}{*}{$>2$} & LO &\multicolumn{6}{c}{359.89}  \\
		& & NLO & 228.31 & 197.84 & 250.26 & 148.04 & 202.64 & 233.92 \\
		& & $K$ & 0.63 & 0.55 & 0.70 & 0.41 & 0.56 & 0.65 \\
		\midrule
		\multirow{6}{*}{\begin{tabular}{@{}l@{}}Ca--Ca \\ ($7.0$~TeV)\end{tabular}} & \multirow{3}{*}{$>0$} & LO&\multicolumn{6}{c}{7320.49}  \\
		& & NLO & 5919.06 & 5468.39 & 6175.74 & 5137.47 & 5737.72 & 6067.01 \\
		& & $K$ & 0.81 & 0.75 & 0.84 & 0.70 & 0.78 & 0.83 \\
		\cmidrule(lr){2-9}
		& \multirow{3}{*}{$>2$} & LO &\multicolumn{6}{c}{581.08 } \\
		& & NLO & 365.55 & 316.96 & 404.19 & 247.32 & 325.47 & 375.81 \\
		& & $K$ & 0.63 & 0.55 & 0.70 & 0.43 & 0.56 & 0.65 \\
		\midrule
		\multirow{6}{*}{{\begin{tabular}{@{}l@{}}O--O \\ ($7.0$~TeV)\end{tabular}}} & \multirow{3}{*}{$>0$} & LO &\multicolumn{6}{c}{ 231.61  }\\
		& & NLO & 187.55 & 172.35 & 195.66 & 163.44 & 181.21 & 191.82 \\
		& & $K$ & 0.81 & 0.74 & 0.84 & 0.71 & 0.78 & 0.83 \\
		\cmidrule(lr){2-9}
		& \multirow{3}{*}{$>2$} & LO & \multicolumn{6}{c}{18.95}  \\
		& & NLO & 12.01 & 10.47 & 13.20 & 7.99 & 10.68 & 12.38 \\
		& & $K$ & 0.63 & 0.55 & 0.70 & 0.42 & 0.56 & 0.65 \\
		\bottomrule
	\end{tabular}
\end{table}

For the A--A systems considered here (O, Ar, Ca, Kr, Xe, Pb), we adopt the quartic charge factor ratios relative to oxygen from Ref.~\cite{Jiang:2024vuq}:
\[
\{Z_{\rm Pb}^4/Z_{\rm Xe}^4,\, Z_{\rm Xe}^4/Z_{\rm Kr}^4,\, Z_{\rm Kr}^4/Z_{\rm Ca}^4,\, Z_{\rm Ca}^4/Z_{\rm Ar}^4,\, Z_{\rm Ar}^4/Z_{\rm O}^4\}
= \{5.10,\, 5.06,\, 10.50,\, 1.52,\, 25.63\}.
\]

Using the LO cross sections at $p_T>0$, the corresponding adjacent-nucleus ratios are
\[
R_{\rm LO}^{AA} = \{\sigma_{\rm LO}^{\rm PbPb/XeXe},\, \sigma_{\rm LO}^{\rm XeXe/KrKr},\, \sigma_{\rm LO}^{\rm KrKr/CaCa},\, \sigma_{\rm LO}^{\rm CaCa/ArAr},\, \sigma_{\rm LO}^{\rm ArAr/OO}\} = \{3.90,\, 3.99,\, 8.52,\, 1.60,\, 19.69\},
\]
while at NLO
\[
R_{\rm NLO}^{AA} = \{\sigma_{\rm NLO}^{\rm PbPb/XeXe},\, \sigma_{\rm NLO}^{\rm XeXe/KrKr},\, \sigma_{\rm NLO}^{\rm KrKr/CaCa},\, \sigma_{\rm NLO}^{\rm CaCa/ArAr},\, \sigma_{\rm NLO}^{\rm ArAr/OO}\} = \{4.18,\, 4.02,\, 8.50,\, 1.61,\, 19.65\}.
\] 
The close agreement between LO and NLO ratios indicates that QCD corrections primarily act as an overall normalization factor and do not alter the nuclear-charge hierarchy of the cross sections.  

A similar but weaker hierarchy is observed in $p$--A collisions. In this case, the coherent enhancement arises from a single nuclear photon source, leading to approximate $Z^2$ scaling, $\sigma_{pA} \propto Z^2$. 
The adjacent-nucleus charge-squared ratios are 
$\{ Z_{\mathrm{Pb}}^2/Z_{\mathrm{Xe}}^2,\, Z_{\mathrm{Xe}}^2/Z_{\mathrm{Kr}}^2,\, Z_{\mathrm{Kr}}^2/Z_{\mathrm{Ca}}^2,$
$Z_{\mathrm{Ca}}^2/Z_{\mathrm{Ar}}^2,\, Z_{\mathrm{Ar}}^2/Z_{\mathrm{O}}^2 \} = \{2.30,\, 2.25,\, 3.24,\, 1.23,\, 5.06\}$, 
in good agreement with the corresponding cross-section ratios
\[
R_{\rm LO}^{pA} = \{\sigma_{\rm LO}^{p{\rm Pb}/p{\rm Xe}},\, \sigma_{\rm LO}^{p{\rm Xe}/p{\rm Kr}},\, \sigma_{\rm LO}^{p{\rm Kr}/p{\rm Ca}},\, \sigma_{\rm LO}^{p{\rm Ca}/p{\rm Ar}},\, \sigma_{\rm LO}^{p{\rm Ar}/p{\rm O}}\} = \{2.16,\, 2.09,\, 2.94,\, 1.27,\, 4.51\},
\]
while at NLO
\[
R_{\rm NLO}^{pA} = \{\sigma_{\rm NLO}^{p{\rm Pb}/p{\rm Xe}},\, \sigma_{\rm NLO}^{p{\rm Xe}/p{\rm Kr}},\, \sigma_{\rm NLO}^{p{\rm Kr}/p{\rm Ca}},\, \sigma_{\rm NLO}^{p{\rm Ca}/p{\rm Ar}},\, \sigma_{\rm NLO}^{p{\rm Ar}/p{\rm O}}\} = \{2.15,\, 2.11,\, 2.92,\, 1.27,\, 4.51\}.
\]

The naive $Z^4$ scaling in A--A collisions is modified when realistic impact-parameter constraints and nuclear form factors are taken into account within the EPA. In particular, the geometric cutoff $b_{\min}\sim 2R_A$ and the kinematic upper limit $E_{\max}\sim \gamma/R_A$ reduce the effective photon luminosity in heavy nuclei, leading to a reduced enhancement relative to the simple $Z^4$ expectation~\cite{Budnev:1975poe,Baur:2001jj,Baltz:2007kq,Klein:2020fmr}. Nevertheless, the ratios between neighboring nuclei remain consistent with the expected $Z^4$ hierarchy.

A similar but less pronounced hierarchy appears in $p$--A collisions, where the coherent enhancement arises from a single nuclear photon source, resulting in approximate $Z^2$ scaling ($\sigma_{pA}\propto Z^2$) and adjacent-nucleus ratios consistent with the charge-squared expectation. In both A--A and $p$--A systems, the deviations from the naive scaling are primarily due to nuclear form factors and impact-parameter constraints, while NLO corrections preserve the same relative hierarchy. This confirms that the system dependence is driven predominantly by the electromagnetic photon flux, rather than by perturbative QCD effects.

This behavior is consistent across all collision systems from $pp$ to heavy-ion UPCs at the HL-LHC and FCC. In all cases, the dominant contribution to the production rate is the coherent photon flux, while NLO QCD corrections mainly rescale the overall normalization ($K<1$) without altering the hierarchy among different systems. This universality suggests that perturbative QCD dynamics factorize to a good approximation from the nuclear electromagnetic structure in exclusive $J/\psi+\gamma$ production.  

\begin{figure}[htbp]
	\centering
	\includegraphics[height=16cm,width=17cm]{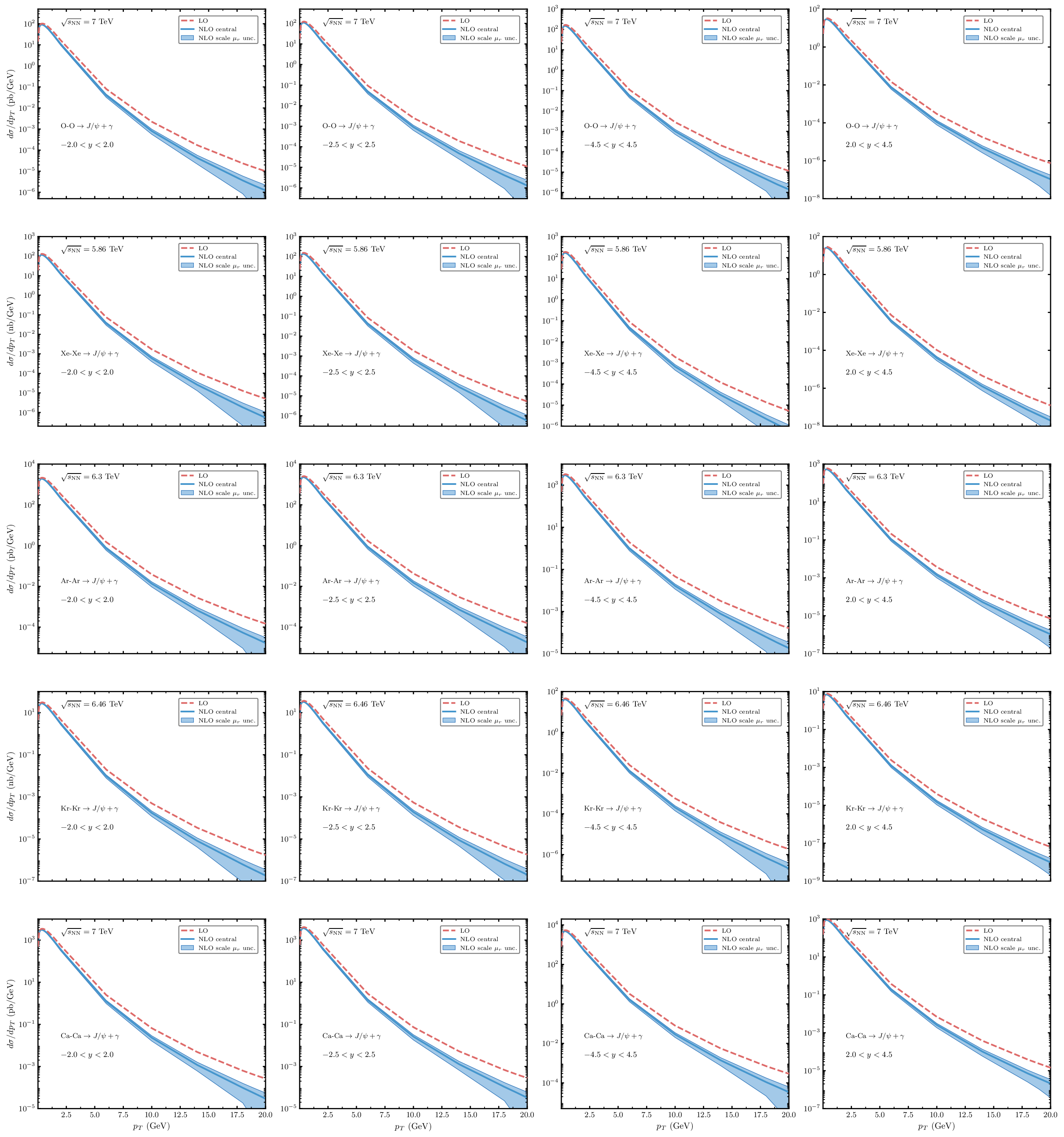}
	\caption{Differential cross sections  $\mathrm{d}\sigma / \mathrm{d}p_T$ in different rapidity ranges for $J/\psi$ production in A--A UPCs at LHC. The dashed line represents LO results, and the shaded band represents NLO results with the renormalization scale varied between $\mu_r/2$ and $2\mu_r$. The solid line in the band indicates the central scale $\mu_r = \sqrt{x_1 x_2 s}$.}
	\label{AA-pt}
\end{figure}

\begin{figure}[htbp]
	\centering
	\includegraphics[height=20cm,width=15cm]{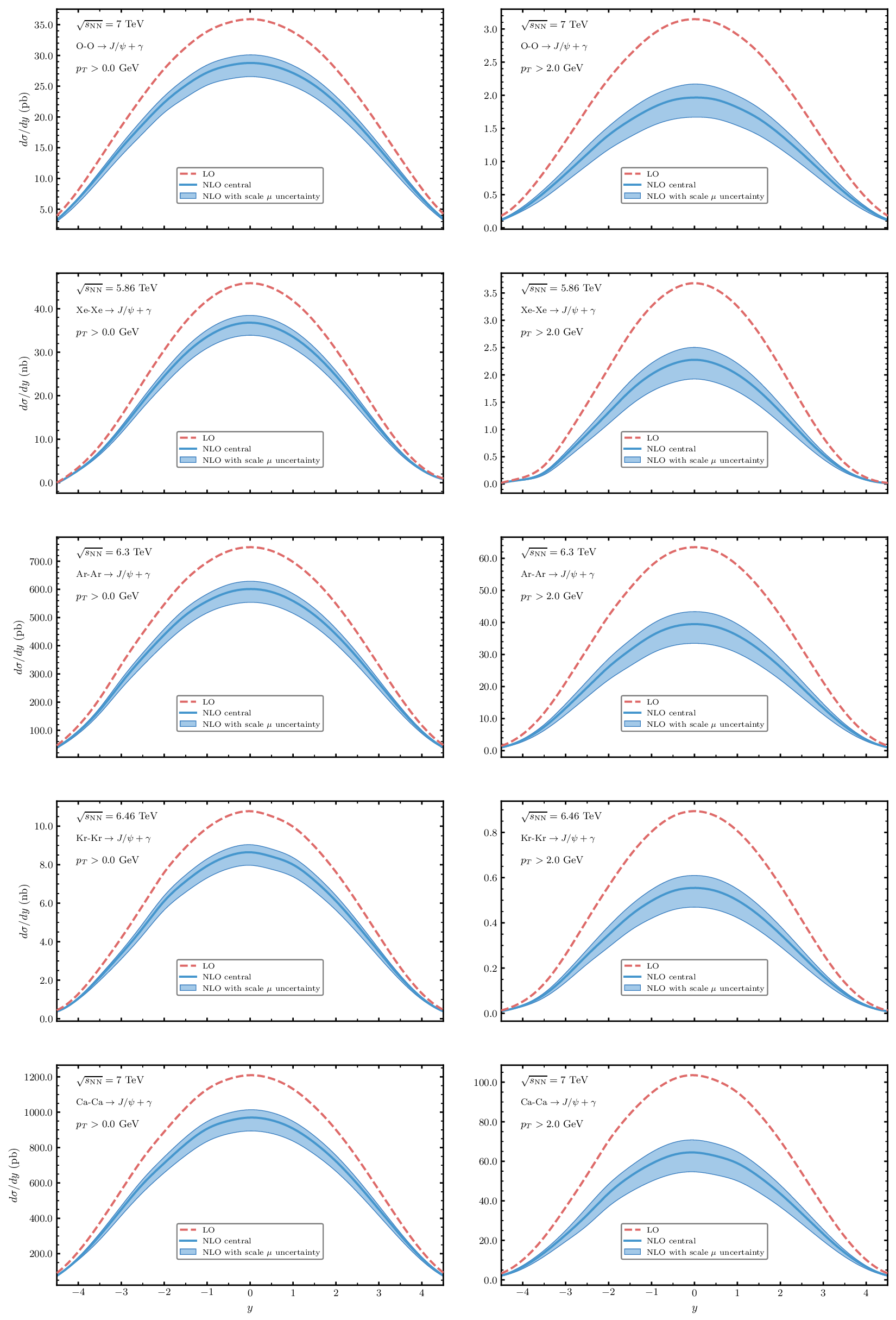}
	\caption{Rapidity distributions $\mathrm{d}\sigma/\mathrm{d}y$  for $J/\psi$ production in A--A UPCs at LHC. The dashed line represents LO results, and the shaded band represents NLO results obtained by varying the renormalization scale between $\mu_r/2$ and $2\mu_r$. The solid line indicates the central scale $\mu_r = \sqrt{x_1 x_2 s}$.}
	\label{AA-y}
\end{figure}

Finally, the transverse-momentum and rapidity distributions of the produced $J/\psi$ in A--A UPCs reflect the generic features of photon–photon fusion in symmetric ion systems. The transverse-momentum distributions, shown in Fig.~\ref{AA-pt}, are primarily determined by the nuclear form factors, with the LO results indicated by the dashed lines and the NLO results shown as shaded bands obtained by varying the renormalization scale between $\mu_r/2$ and $2\mu_r$. The central value corresponds to $\mu_r = \sqrt{x_1 x_2 s}$. The rapidity distributions, shown in Fig.~\ref{AA-y}, are symmetric around $y=0$, and the NLO corrections preserve the same overall pattern as the LO results. In $p$--A collisions, the asymmetry of the photon flux between the proton and the nucleus shifts the rapidity distributions, while the $p_T$ dependence and the impact of QCD corrections remain similar to those in A--A systems. This indicates that the essential production dynamics are already captured by the A--A results, with $p$--A systems differing mainly due to the asymmetric photon flux.

\subsection{Expected event yields at HL-LHC and FCC}\label{sec3.4}
To assess the experimental observability of the process, we estimate the expected event yields using  
\begin{equation}
	N = \sigma_{\rm NLO}\,\mathcal{L}_{\rm int}\,
	\mathrm{Br}(J/\psi \to \mu^+\mu^-),
\end{equation}
where $\mathrm{Br}(J/\psi \to \mu^+\mu^-)=5.961\%$~\cite{ParticleDataGroup:2024cfk}. 
Using the NLO cross sections evaluated at the central scale 
$\mu_r=\sqrt{x_1x_2s}$, the expected numbers of 
reconstructible $J/\psi+\gamma$ events for all considered collision 
systems are summarized in Table~\ref{tab:events}.

\begin{table*}[htbp]
	\centering
	\caption{Expected numbers of reconstructible 
		$J/\psi+\gamma$ events via $J/\psi\to\mu^+\mu^-$ at NLO 
		(using the central scale $\mu_r=\sqrt{x_1x_2s}$). 
		Design integrated luminosities are used for each collider mode~\cite{Citron:2018lsq,Bruce:2018yzs,dEnterria:2022sut,FCC:2018vvp,Dainese:2016gch}.}
	\label{tab:events}
	\small
	\renewcommand{\arraystretch}{1.0} 
	\begin{tabular}{lccccc}
		\toprule
		Collider & System & $\sqrt{s_{NN}}$ (TeV) & $\mathcal{L}_{\rm int}$ & $N(p_T>0)$ & $N(p_T>2)$ \\
		\midrule
		\multirow{8}{*}{HL-LHC}
		& Pb--Pb & 5.52  & $5~\mathrm{nb}^{-1}$ & 245 & 13 \\
		& Xe--Xe & 5.86  & $30~\mathrm{nb}^{-1}$ & 362 & 21 \\
		& Kr--Kr & 6.46  & $120~\mathrm{nb}^{-1}$ & 360 & 22 \\
		& Ar--Ar & 6.3  & $1.1~\mathrm{pb}^{-1}$ & 242 & 15 \\
		& Ca--Ca & 7.0  & $0.8~\mathrm{pb}^{-1}$ & 282 & 17 \\
		& O--O   & 7.0  & $12~\mathrm{pb}^{-1}$ & 134 & 9 \\
		& p--Pb  & 8.8  & $1~\mathrm{pb}^{-1}$ & 22 & 1 \\
		& pp     & 14  & $150~\mathrm{fb}^{-1}$ & $1.1\times10^3$ & $84$ \\
		\midrule
		\multirow{3}{*}{FCC}
		& Pb--Pb & 39.4  & $110~\mathrm{nb}^{-1}$ & $2.0\times10^4$ & $1.4\times10^3$ \\
		& p--Pb  & 62.8 & $29~\mathrm{pb}^{-1}$  & $1.5\times10^3$ & $1.2\times10^2$ \\
		& pp     & 100   & $1~\mathrm{ab}^{-1}$   & $1.4\times10^4$ & $1.1\times10^3$ \\
		\bottomrule
	\end{tabular}
\end{table*}

Although the $\gamma\gamma$ production cross section grows approximately with the quartic nuclear-charge dependence $\sigma \sim Z^4$, the observable event yield is ultimately determined by the combined effect of this enhancement and the integrated luminosity. At the HL-LHC, medium-mass nucleus systems such as Xe--Xe and Kr--Kr provide the largest statistics, slightly exceeding Pb--Pb despite their smaller nuclear charge, owing to their substantially higher integrated luminosities. Light-nucleus systems (O--O, Ar--Ar, Ca--Ca) remain experimentally accessible, while the $p$--Pb and $pp$ channels are primarily luminosity driven and yield 
smaller samples.

The situation is markedly different at the FCC. The higher collision energies significantly enhance the photon luminosity, while the integrated luminosities are substantially larger than those at the HL-LHC. As a consequence, the process becomes measurable with high statistics in all collision systems. In particular, pp and Pb--Pb 
collisions can yield up to $\mathcal{O}(10^{3}\text{--}10^{4})$ reconstructible events. The $p$--Pb systems also provides sizable event samples, reaching $\mathcal{O}(10^{2}\text{--}10^{3})$ events.

Overall, the hierarchy of event yields across collision systems is controlled primarily by the electromagnetic photon flux, while QCD corrections mainly affect the overall normalization. In this respect, the FCC offers a favorable setting for detailed studies of exclusive $J/\psi+\gamma$ production. At the HL-LHC, the process remains within reach of exploratory measurements, especially in medium-mass ion modes where the achievable statistics are comparatively higher.

\section{Conclusion}\label{sec4}

In this work we have performed a comprehensive NRQCD study of exclusive 
$\gamma\gamma \to J/\psi+\gamma$ production in UPCs, including the complete NLO QCD corrections and impact-parameter dependent survival effects. Predictions have been obtained for $pp$, $p$--Pb, and a broad set of nucleus--nucleus collision systems at HL-LHC and FCC energies.

At the level of total cross sections, the LO results exhibit the expected electromagnetic hierarchy dictated by the nuclear charge, $\sigma_{AA} \gg \sigma_{pA} \gg \sigma_{pp}$. The NLO QCD corrections primarily rescale the cross sections, reducing them by roughly $20\%$ ($K\sim 0.8$), while stronger suppressions occur when a transverse-momentum cut is applied. This behavior indicates that $p_T$ selections enhance the sensitivity to virtual corrections by probing different kinematic regions.

The production cross sections approximately follow a $Z^4$ scaling for A--A collisions and a $Z^2$ scaling for $p$--A collisions. The rapidity distributions, illustrated for A--A collisions are symmetric around $y=0$ due to the comparable photon fluxes from the two nuclei. In contrast, $p$--A collisions feature asymmetric photon fluxes, shifting the rapidity spectra, as already seen in the $p$--Pb results. 
Despite this asymmetry, the $p_T$ dependence and the relative effect of QCD corrections remain similar to those in A--A systems, indicating that the underlying production dynamics are largely captured by the nuclear--nuclear results, with the proton--nucleus systems differing mainly because of the asymmetric photon flux.

Although the $Z^4$ scaling suggests that the heaviest nuclei should yield the largest event rates, the actual statistics depend on the interplay between electromagnetic enhancement and the integrated luminosities. 
At the HL-LHC, medium-mass systems such as Xe--Xe and Kr--Kr can provide larger event samples than Pb--Pb due to their substantially higher luminosities, 
while light-nucleus systems (O--O, Ar--Ar, Ca--Ca) are still experimentally accessible. 
The $pp$ and $p$--Pb channels, by contrast, are largely luminosity limited and produce smaller event samples.

At the FCC, higher collision energies increase the photon luminosity, and the 
projected integrated luminosities are orders of magnitude larger than at the 
HL-LHC. Consequently, exclusive $J/\psi+\gamma$ production becomes statistically 
robust in all collision systems. In particular, $pp$ and Pb--Pb collisions may yield up to 
$\mathcal{O}(10^{3}\text{--}10^{4})$ reconstructible events, while significant samples are also expected in a $p$--Pb system.

A moderate dependence on the renormalization scale persists at NLO, indicating that higher-order corrections would be needed to further stabilize 
the predictions \cite{Sang:2023liy,Huang:2022dfw}. Nevertheless, the present 
analysis establishes a consistent framework for describing exclusive quarkonium–
photon production in UPCs, offering concrete reference values for future measurements. The same approach can be extended to other nuclear systems and two-photon scattering processes.

\begin{acknowledgments}
We would like to express our sincere gratitude to Professor Zhi-Guo He for invaluable guidance and insightful discussions regarding our research work. This work was supported by the National Natural Science Foundation of China (Grants No.12575087, No.11705078).
\end{acknowledgments}

\subsection*{DATA AVAILABILITY}
\parbox{\linewidth}{Numerical results including cross sections and differential kinematic distributions presented in this paper are publicly available in the Zenodo repository~(https://doi.org/10.5281/zenodo.20574659). The original numerical codes used for calculations can be obtained from the authors upon reasonable request.}


\begin{thebibliography}{99}
\bibitem{LHCb:2015foc}
R.~Aaij \textit{et al.} [LHCb],
Measurement of forward $J/\psi$ production cross sections in $pp$ collisions at $\sqrt{s}=13$ TeV,
JHEP \textbf{10}, 172 (2015)

\bibitem{CDF:2009kwm}
T.~Aaltonen \textit{et al.} [CDF], Production of psi(2S) Mesons in p anti-p Collisions at 1.96-TeV,
Phys. Rev. D \textbf{80}, 031103 (2009)

\bibitem{PHENIX:2009ghc}
A.~Adare \textit{et al.} [PHENIX], Transverse momentum dependence of J/psi polarization at midrapidity in p+p collisions at $\sqrt{s} = 200~ \text{GeV}$,
Phys. Rev. D \textbf{82}, 012001 (2010)

\bibitem{Faccioli:2022alj}
P.~Faccioli and C.~Louren{\c{c}}o, On the polarization of the non-prompt contribution to inclusive J/{\ensuremath{\psi}} production in pp collisions, JHEP \textbf{10}, 005 (2022)

\bibitem{Xu:2012am}
G.~Z.~Xu, Y.~J.~Li, K.~Y.~Liu and Y.~J.~Zhang,
Relativistic Correction to Color Octet J/psi Production at Hadron Colliders,
Phys. Rev. D \textbf{86}, 094017 (2012)

\bibitem{Li:2019mdx}
R.~Li, A.~P.~Chen, J.~K.~Huang and Y.~Q.~Ma, Relativistic effect of $J/\psi$ hadroproduction in large p$_{T}$ region,
JHEP \textbf{10}, 210 (2019)

\bibitem{Bodwin:1994jh}
G.~T.~Bodwin, E.~Braaten and G.~P.~Lepage,
Rigorous QCD analysis of inclusive annihilation and production of heavy quarkonium,
Phys. Rev. D \textbf{51}, 1125 (1995)

\bibitem{Baltz:2007kq}
A.~J.~Baltz \textit{et al.},
The Physics of Ultraperipheral Collisions at the LHC,
Phys. Rept. \textbf{458}, 1 (2008)

\bibitem{Bertulani:2005ru}
C.~A.~Bertulani, S.~R.~Klein and J.~Nystrand,
Physics of ultra-peripheral nuclear collisions,
Ann. Rev. Nucl. Part. Sci. \textbf{55}, 271 (2005)

\bibitem{Klein:2019qfb}
S.~R.~Klein and H.~M{\"a}ntysaari,
Imaging the nucleus with high-energy photons,
Nature Rev. Phys. \textbf{1}, 662 (2019)

\bibitem{Lansberg:2024zap}
J.~P.~Lansberg, K.~Lynch, C.~Van Hulse and R.~McNulty,
Inclusive photoproduction of vector quarkonium in ultra-peripheral collisions at the LHC,
Eur. Phys. J. C \textbf{85}, 161 (2025)

\bibitem{Klein:2020fmr}
S.~Klein and P.~Steinberg,
Photonuclear and Two-photon Interactions at High-Energy Nuclear Colliders,
Ann. Rev. Nucl. Part. Sci. \textbf{70}, 323 (2020)

\bibitem{Brodsky:1971ud}
S.~J.~Brodsky, T.~Kinoshita and H.~Terazawa,
Two Photon Mechanism of Particle Production by High-Energy Colliding Beams,
Phys. Rev. D \textbf{4}, 1532 (1971)

\bibitem{Budnev:1975poe}
V.~M.~Budnev, I.~F.~Ginzburg, G.~V.~Meledin and V.~G.~Serbo,
The Two photon particle production mechanism. Physical problems. Applications. Equivalent photon approximation,
Phys. Rept. \textbf{15}, 181 (1975)

\bibitem{LHCb:2013nqs}
R.~Aaij \textit{et al.} [LHCb],
Exclusive $J/\psi$ and $\psi$(2S) production in pp collisions at $ \sqrt{s} = 7$ TeV,
J. Phys. G \textbf{40}, 045001 (2013)

\bibitem{LHCb:2014acg}
R.~Aaij \textit{et al.} [LHCb],
Updated measurements of exclusive $J/\psi$ and $\psi$(2S) production cross sections in pp collisions at $\sqrt{s}=7$ TeV,
J. Phys. G \textbf{41}, 055002 (2014)

\bibitem{LHCb:2016oce}
[LHCb],
Central exclusive production of $J/\psi$ and $\psi(2S)$ mesons in pp collisions at $\sqrt{s}=13$ TeV, LHCb-CONF-2016-007 (2016)

\bibitem{ALICE:2013wjo}
E.~Abbas \textit{et al.} [ALICE],
Charmonium and $e^+e^-$ pair photoproduction at mid-rapidity in ultra-peripheral Pb-Pb collisions at $\sqrt{s_{\rm NN}}$=2.76 TeV,
Eur. Phys. J. C \textbf{73}, 2617 (2013)

\bibitem{ALICE:2018oyo}
S.~Acharya \textit{et al.} [ALICE],
Energy dependence of exclusive $\mathrm {J}/\psi $ photoproduction off protons in ultra-peripheral p{\textendash}Pb collisions at $\sqrt{s_{\mathrm {\scriptscriptstyle NN}}} = 5.02$ TeV,
Eur. Phys. J. C \textbf{79}, 402 (2019)

\bibitem{ALICE:2023mfc}
S.~Acharya \textit{et al.} [ALICE],
Exclusive and dissociative J/{\ensuremath{\psi}} photoproduction, and exclusive dimuon production, in p-Pb collisions at sNN=8.16{\,}{\,}TeV,
Phys. Rev. D \textbf{108}, 112004 (2023)

\bibitem{ALICE:2023kgv}
S.~Acharya \textit{et al.} [ALICE],
Photoproduction of K+K- Pairs in Ultraperipheral Collisions,
Phys. Rev. Lett. \textbf{132}, 222303 (2024)

\bibitem{PHENIX:2009xtn}
S.~Afanasiev \textit{et al.} [PHENIX],
Photoproduction of J/psi and of high mass e+e- in ultra-peripheral Au+Au collisions at $\sqrt{s}$  = 200 GeV,
Phys. Lett. B \textbf{679}, 321 (2009)

\bibitem{Rapp:2008tf}
R.~Rapp, D.~Blaschke and P.~Crochet,
Charmonium and bottomonium production in heavy-ion collisions,
Prog. Part. Nucl. Phys. \textbf{65}, 209 (2010)

\bibitem{Klasen:2008mh}
M.~Klasen and J.~P.~Lansberg,
Perspectives for inclusive quarkonium production in photon-photon collisions at the LHC, Nucl. Phys. B Proc. Suppl. \textbf{179-180}, 226 (2008)

\bibitem{Zheng:2024mep}
S.~Zheng, J.~Liu, S.~Zheng, J.~Zhao and B.~Chen,
Charmonia production in hot QCD matter and electromagnetic fields,
Phys. Rev. C \textbf{111}, 054908 (2025)

\bibitem{Jiang:2024vuq}
J.~Jiang, S.~Y.~Li, X.~Liang, Y.~R.~Liu, C.~F.~Qiao, Z.~G.~Si and H.~Yang,
Pseudoscalar heavy quarkonium production in heavy ion ultraperipheral collision,
Phys. Rev. D \textbf{110}, 054005 (2024)

\bibitem{Obertova:2024nmb}
J.~{\'O}bertov{\'a} and J.~Nemchik,
Path integral treatment of coherence effects in charmonium production in nuclear ultra-peripheral collisions,
PoS \textbf{ICHEP2024}, 505 (2025)

\bibitem{Azevedo:2024bqd}
C.~N.~Azevedo, F.~C.~Sobrinho and F.~S.~Navarra,
Production of $\eta_b$ in ultra-peripheral $Pb Pb$ collisions,
arXiv:2412.18567 [hep-ph]

\bibitem{Klasen:2004az}
M.~Klasen, B.~A.~Kniehl, L.~N.~Mihaila and M.~Steinhauser,
$J/\psi$ plus prompt-photon associated production in two-photon collisions at next-to-leading order,
Phys. Rev. D \textbf{71}, 014016 (2005)

\bibitem{Goncalves:2023sts}
V.~P.~Goncalves, M.~Klasen and B.~D.~Moreira,
Exclusive $J/\Psi$ plus jet associated production in ultraperipheral $PbPb$ collisions,
Eur. Phys. J. C \textbf{83}, 895 (2023)

\bibitem{Chen:2025gwp}
Z.~Q.~Chen and L.~B.~Chen,
Exclusive J/{\ensuremath{\psi}}+{\ensuremath{\gamma}} production in ultraperipheral ion collisions,
Phys. Rev. D \textbf{112}, 056013 (2025)

\bibitem{Shao:2022cly}
H.~S.~Shao and D.~d'Enterria,
gamma-UPC: automated generation of exclusive photon-photon processes in ultraperipheral proton and nuclear collisions with varying form factors,
JHEP \textbf{09}, 248 (2022)

\bibitem{Eichten:1995ch}
E.~J.~Eichten and C.~Quigg,
Quarkonium wave functions at the origin,
Phys. Rev. D \textbf{52}, 1726 (1995)

\bibitem{Cahn:1990jk}
R.~N.~Cahn and J.~D.~Jackson,
Realistic equivalent photon yields in heavy ion collisions,
Phys. Rev. D \textbf{42}, 3690 (1990)

\bibitem{Glauber:1970jm}
R.~J.~Glauber and G.~Matthiae,
High-energy scattering of protons by nuclei,
Nucl. Phys. B \textbf{21}, 135 (1970)

\bibitem{dEnterria:2020dwq}
D.~d'Enterria and C.~Loizides,
Progress in the Glauber Model at Collider Energies,
Ann. Rev. Nucl. Part. Sci. \textbf{71}, 315 (2021)

\bibitem{Loizides:2017ack}
C.~Loizides, J.~Kamin and D.~d'Enterria,
Improved Monte Carlo Glauber predictions at present and future nuclear colliders,
Phys. Rev. C \textbf{97}, 054910 (2018); [erratum: Phys. Rev. C \textbf{99}, 019901 (2019)]

\bibitem{Loizides:2014vua}
C.~Loizides, J.~Nagle and P.~Steinberg,
Improved version of the PHOBOS Glauber Monte Carlo,
SoftwareX \textbf{1-2}, 13 (2015)

\bibitem{Frankfurt:2006jp}
L.~Frankfurt, C.~E.~Hyde, M.~Strikman and C.~Weiss,
Generalized parton distributions and rapidity gap survival in exclusive diffractive $p p$ scattering,
Phys. Rev. D \textbf{75}, 054009 (2007)

\bibitem{He:2024lrb}
Z.~G.~He, X.~B.~Jin, B.~A.~Kniehl and R.~Li,
Next-to-leading-order relativistic and QCD corrections to prompt pair photoproduction at future colliders,
Chin. Phys. C \textbf{48}, 083107 (2024)

\bibitem{Nogueira:1991ex}
P.~Nogueira,
Automatic Feynman Graph Generation,
J. Comput. Phys. \textbf{105}, 279 (1993)

\bibitem{Vermaseren:2000nd}
J.~A.~M.~Vermaseren,
New features of FORM,
arXiv:math-ph/0010025

\bibitem{vonManteuffel:2012np}
A.~von Manteuffel and C.~Studerus,
Reduze 2 - Distributed Feynman Integral Reduction,
arXiv:1201.4330 [hep-ph]

\bibitem{Smirnov:2019qkx}
A.~V.~Smirnov and F.~S.~Chukharev,
FIRE6: Feynman Integral REduction with modular arithmetic,
Comput. Phys. Commun. \textbf{247}, 106877 (2020)

\bibitem{Patel:2016fam}
H.~H.~Patel,
Package-X 2.0: A Mathematica package for the analytic calculation of one-loop integrals,
Comput. Phys. Commun. \textbf{218}, 66 (2017)

\bibitem{Ellis:2007qk}
R.~K.~Ellis and G.~Zanderighi,
Scalar one-loop integrals for QCD,
JHEP \textbf{02}, 002 (2008)

\bibitem{Hahn:2004fe}
T.~Hahn,
CUBA: A Library for multidimensional numerical integration,
Comput. Phys. Commun. \textbf{168}, 78 (2005)

\bibitem{Hahn:2014fua}
T.~Hahn,
Concurrent Cuba,
J. Phys. Conf. Ser. \textbf{608}, 012066 (2015)

\bibitem{DeJager:1974liz}
C.~W.~De Jager, H.~De Vries and C.~De Vries,
Nuclear charge and magnetization density distribution parameters from elastic electron scattering,
Atom. Data Nucl. Data Tabl. \textbf{14}, 479 (1974); [erratum: Atom. Data Nucl. Data Tabl. \textbf{16}, 580 (1975)]

\bibitem{DeVries:1987atn}
H.~De Vries, C.~W.~De Jager and C.~De Vries,
Nuclear charge and magnetization density distribution parameters from elastic electron scattering,
Atom. Data Nucl. Data Tabl. \textbf{36}, 495 (1987)

\bibitem{Baur:2001jj}
G.~Baur, K.~Hencken, D.~Trautmann, S.~Sadovsky and Y.~Kharlov,
Coherent gamma gamma and gamma-A interactions in very peripheral collisions at relativistic ion colliders,
Phys. Rept. \textbf{364}, 359 (2002)

\bibitem{Liu:2003jj}
K.~Y.~Liu, Z.~G.~He and K.~T.~Chao,
Inclusive charmonium production via double $c \bar{c}$ in $e^{+} e^{-}$ annihilation,
Phys. Rev. D \textbf{69}, 094027 (2004)

\bibitem{Campbell:2007ws}
J.~M.~Campbell, F.~Maltoni and F.~Tramontano,
QCD corrections to J/psi and Upsilon production at hadron colliders,
Phys. Rev. Lett. \textbf{98}, 252002 (2007)

\bibitem{Gong:2008sn}
B.~Gong and J.~X.~Wang,
Next-to-leading-order QCD corrections to $J/\psi$ polarization at Tevatron and Large-Hadron-Collider energies,
Phys. Rev. Lett. \textbf{100}, 232001 (2008)

\bibitem{Shao:2014yta}
H.~S.~Shao, H.~Han, Y.~Q.~Ma, C.~Meng, Y.~J.~Zhang and K.~T.~Chao,
Yields and polarizations of prompt $J/\psi$ and $\psi(2S)$ production in hadronic collisions,
JHEP \textbf{05}, 103 (2015)

\bibitem{Chao:2012iv}
K.~T.~Chao, Y.~Q.~Ma, H.~S.~Shao, K.~Wang and Y.~J.~Zhang,
$J/\psi$ Polarization at Hadron Colliders in Nonrelativistic QCD,
Phys. Rev. Lett. \textbf{108}, 242004 (2012)

\bibitem{Ma:2010jj}
Y.~Q.~Ma, K.~Wang and K.~T.~Chao,
A complete NLO calculation of the $J/\psi$ and $\psi'$ production at hadron colliders,
Phys. Rev. D \textbf{84}, 114001 (2011)

\bibitem{Pumplin:2002vw}
J.~Pumplin, D.~R.~Stump, J.~Huston, H.~L.~Lai, P.~M.~Nadolsky and W.~K.~Tung,
New generation of parton distributions with uncertainties from global QCD analysis,
JHEP \textbf{07}, 012 (2002)

\bibitem{Goncalves:2018yxc}
V.~P.~Gon{\c{c}}alves and B.~D.~Moreira,
$\eta_c$ production in photon-induced interactions at the LHC,
Phys. Rev. D \textbf{97}, 094009 (2018)

\bibitem{Citron:2018lsq}
Z.~Citron \textit{et al.},
Report from Working Group 5: Future physics opportunities for high-density QCD at the LHC with heavy-ion and proton beams,
CERN Yellow Rep. Monogr. \textbf{7}, 1159 (2019)

\bibitem{ALICE:2014eof}
B.~B.~Abelev \textit{et al.} [ALICE],
Exclusive $\mathrm{J/}\psi$ photoproduction off protons in ultra-peripheral p-Pb collisions at $\sqrt{s_{\rm NN}}=5.02$ TeV,
Phys. Rev. Lett. \textbf{113}, 232504 (2014)

\bibitem{Xu:1998rp}
J.~S.~Xu and H.~A.~Peng,
Associated $J/\psi$ + $\gamma$ production through double pomeron exchange:
The Nature of the pomeron and hard diffractive factorization breaking,''
Commun. Theor. Phys. \textbf{34}, 129-134 (2000)

\bibitem{GayDucati:2009rr}
M.~B.~Gay Ducati, M.~M.~Machado and M.~V.~T.~Machado,
Diffractive quarkonium production in association with a photon at the LHC,''
Phys. Lett. B \textbf{683}, 150-153 (2010)

\bibitem{Goncalves:2012bt}
V.~P.~Goncalves and M.~M.~Machado,
Quarkonium+$\gamma$ production in coherent hadron - hadron interactions at
LHC energies,
Eur. Phys. J. C \textbf{72}, 2231 (2012)

\bibitem{Lebiedowicz:2023mhe}
P.~Lebiedowicz, O.~Nachtmann and A.~Szczurek,
Central exclusive diffractive production of a single photon in high-energy proton-proton collisions within the tensor-Pomeron approach,
Phys. Rev. D \textbf{107}, no.7, 074014 (2023)

\bibitem{ParticleDataGroup:2024cfk}
S.~Navas \textit{et al.} [Particle Data Group],
Review of particle physics,
Phys. Rev. D \textbf{110}, 030001 (2024)

\bibitem{Bruce:2018yzs}
R.~Bruce \textit{et al.},
New physics searches with heavy-ion collisions at the CERN Large Hadron Collider,
J. Phys. G \textbf{47}, 060501 (2020)

\bibitem{dEnterria:2022sut}
D.~d'Enterria \textit{et al.},
Opportunities for new physics searches with heavy ions at colliders,
J. Phys. G \textbf{50}, 050501 (2023)

\bibitem{FCC:2018vvp}
A.~Abada \textit{et al.} [FCC],
FCC-hh: The Hadron Collider: Future Circular Collider Conceptual Design Report Volume 3,
Eur. Phys. J. ST \textbf{228}, 755 (2019)

\bibitem{Dainese:2016gch}
A.~Dainese \textit{et al.},
Heavy ions at the Future Circular Collider,
CERN Yellow Rep. 3, 635 (2017)

\bibitem{Sang:2023liy}
W.~L.~Sang, F.~Feng, Y.~Jia, Z.~Mo, J.~Pan and J.~Y.~Zhang,
Optimized O({\ensuremath{\alpha}}s2) Correction to Exclusive Double-J/{\ensuremath{\psi}} Production at B Factories,
Phys. Rev. Lett. \textbf{131}, 161904 (2023)

\bibitem{Huang:2022dfw}
X.~D.~Huang, B.~Gong and J.~X.~Wang,
Next-to-next-to-leading-order QCD corrections to J/{\ensuremath{\psi}} plus {\ensuremath{\eta}}$_{c}$ production at the B factories,
JHEP \textbf{02}, 049 (2023)
\end{thebibliography}

\end{document}